\documentclass{aastex61}
\usepackage{amsmath}

\def\Msol{M_\odot}

\received{June 1, 2018}
\revised{June 27, 2018}
\accepted{\today}
\submitjournal{ApJ}

\shorttitle{Supernovae with JWST}
\shortauthors{Reg\H os \& Vink\'o}

\begin{document}

\title{Detection and classification of supernovae beyond $z \sim 2$ redshift with the {\it James Webb Space Telescope} }

\correspondingauthor{Jozsef Vinko}
\email{vinko@konkoly.hu}

\author[0000-0002-9498-4957]{Enik\H o Reg\H os}
\affil{ Konkoly Observatory, MTA CSFK, 
Konkoly-Thege M. ut 15-17,
Budapest, 1121, Hungary}

\author[0000-0001-8764-7832]{J\'ozsef Vink\'o}
\affil{ Konkoly Observatory, MTA CSFK, Konkoly-Thege M. ut 15-17, 
Budapest, 1121, Hungary}
\affil{Department of Optics \& Quantum Electronics, University of Szeged, D\'om t\'er 9, Szeged, 6720, Hungary}

\begin{abstract}
Future time-domain surveys for transient events in the near- and mid-infrared bands will significantly extend our understanding about the physics of the early Universe. In this paper we study the implications of a deep ($\sim 27$ mag), long-term ($\sim 3$ years), observationally inexpensive survey with the James Webb Space Telescope ({\it JWST}) within its Continuous Viewing Zone, aimed at discovering luminous supernovae beyond $z \sim 2$ redshift. We explore the possibilities for detecting Superluminous Supernovae (SLSNe) as well as Type Ia supernovae at such high redshifts and estimate their expected numbers within a relatively small ($\sim 0.1$ deg$^2$) survey area. It is found that we can expect $\sim 10$ new SLSNe and $\sim 50$ SNe Ia discovered in the $1 < z < 4$ redshift range. We show that it is possible to get relatively accurate ($\sigma_z \lesssim 0.25$) photometric redshifts for Type Ia SNe by fitting their Spectral Energy Distributions (SED), redshifted into the observed near-IR bands, with SN templates. We propose that Type Ia SNe occupy a relatively narrow range on the {\it JWST} F220W$-$F440W vs F150W$-$F356W color-color diagram between $\pm 7$ rest-frame days around maximum light, which could be a useful classification tool for such type of transients. We also study the possibility of extending the Hubble-diagram of Type Ia SNe beyond redshift 2 up to $z \sim 4$. Such high-$z$ SNe Ia may provide new observational constraints for their progenitor scenario.

\end{abstract}

%

\section{Introduction}\label{sec:intro}

One of the fundamental questions of modern astrophysics and cosmology is related to the problem of star formation in the early Universe: how did the Universe make its first stars? Decade-long observational and theoretical efforts have been devoted to reveal and establish the cosmic Star Formation Rate (SFR) as a function of redshift \citep[see e.g.][and references therein]{hopkins06, behroozi13, madi14, bouwens14, oesch15}. This function gives the mass of newborn stars per year per volume element, and it is a strong function of the cosmic time, i.e. redshift, up to $z \sim 10$ \citep{oesch18}.

One of the exciting possibilities to probe the cosmic SFR at various redshifts is the discovery of new transients that are related to the death of massive stars, i.e. long-duration gamma-ray bursts (LGRBs) and supernovae (SNe). LGRBs are now routinely detected at
redshifts beyond $z \sim 2$, but it is still very challenging observationally for SNe. Because the upcoming near- and mid-infrared surveys offer new opportunities for such efforts \citep[e.g.][]{tanaka13}, \citet{wang17} proposed the First Lights At REionization (FLARE) project for discovering various types of transients with  NASA's James Webb Space Telescope ({\it JWST}) at the highest possible redshifts.  One of the most important (and most ambitious) goals of the FLARE project is the discovery of the most distant, most luminous supernovae with {\it JWST}. 

Superluminous supernovae (SLSNe), which have the highest intrinsic peak luminosities among SNe known to date, seem to be promising targets for such a purpose, because they can be potentially detected up to $z \sim 10$ redshifts with deep ($m_{AB} \gtrsim 26$ mag) surveys \citep{tanaka13}. As they are produced by very massive progenitors, they can closely trace the cosmic star formation rate variation along redshift. Thus, discovering SLSNe at very high redshifts can provide unprecedented information on the history of early star formation and evolution. 

Thermonuclear (Type Ia) SNe offer another possibility to shed light on star forming processes in the early Universe. Type Ia SNe are fainter, but more abundant (at least in the local Universe) than SLSNe. They have \(M_V \sim -19 \pm 1\) absolute AB magnitude at peak, and relatively UV-faint SED at and after maximum light. They are produced by exploding white dwarfs (WDs): either a single mass-gaining WD near the Chandrasekhar limit (single degenerate channel, SD) or two merging WDs (double-degenerate channel, DD) \citep{maoz14, livio18}.

Because WDs are formed from low-mass ($\lesssim 8$ M$_\odot$) stars at the end of their lifetime, a delay between the birth of a new star and the explosion of the WD is expected \citep[e.g.][]{graur14, maoz14}. The delay-time distribution (DTD) depends on the progenitor channel, i.e. the SD and DD scenarios. A ``prompt'' channel that contains SNe Ia that explode very shortly ($\lesssim 500$ Myr) after the formation of the WD were examined by \citet[e.g.][]{scanna05,raskin09}.
The existence of such a ``prompt'' Ia population can be critically probed with detections of high-$z$ SNe Ia
\citep[][]{regos13, rodney14}. Furthermore, direct measurements of the SN Ia DTD may help in distinguishing between the SD and DD scenarios. It is possible that both channels operate, either on short (SD) or long (DD) time scales. For example, the combined data from the CLASH \citep{postman12} and CANDELS \citep{koeke11,grogin11} surveys are consistent with long delay times corresponding to the DD scenario \citep{rodney14}.


In this paper we focus on one particular topic within the FLARE project: observing  the most distant, most luminous supernovae with {\it JWST}. 
We explore the feasibility of detecting and classifying different types of SNe, thermonuclear (Type Ia SNe) and superluminous supernovae (SLSNe) in particular, beyond $z \sim 2$ with {\it JWST}, as well as measuring their physical properties from spectrophotometry and extending the observed Hubble-diagram for SNe Ia at as high redshifts as possible. 


High-redshift Type Ia SNe are used to derive cosmological parameters from their Hubble-diagram \citep[see e.g.][and references therein]{scolnic18}. With the help of such SNe it is possible to extend the Hubble diagram to $z \gtrsim 1.5$, probing progenitor evolution separately from the nature of dark energy \citep[e.g.][]{riess18}. To study the
properties of dark energy one measures its equation of state $w$ and time variation to distinguish among cosmological explanations. Departure from $-1$ or detection of $dw/dz$ would indicate a present
epoch of weak inflation.
Detecting SNe Ia at $1.5 < z < 2.5$ provides the unique chance to test SN Ia distance measurements for the deleterious effects of evolution independent of our ignorance of dark energy \citep{riess18}. 
We can also test DD and SD scenarios by measuring the SNe Ia delay time distribution.
The CLASH  and CANDELS  programs provided measurement of the SN Ia rate up to $z \sim 2$ \citep{rodney14}, and FLARE, as planned, will be capable of going beyond 2.

\section{Survey strategy for high-redshift transients using {\it JWST}}\label{sec:survey}

Recently \citet{tanaka13} showed that a moderately deep ($\sim 26$ mag) $\sim 100$ deg$^2$ survey in the near-infrared (NIR) can potentially discover $\sim 10$ SLSNe up to redshift $z \sim 10$. With a $\sim 1$ mag deeper limiting magnitude the redshift limit could be pushed even further, toward $z\sim 15$ \citep{tanaka13}. 

Because such kind of a survey is observationally very expensive with {\it JWST}, \citet{wang17} proposed another observing strategy for discovering high-redshift SNe: continuous monitoring of a smaller area toward the North Ecliptic Pole with {\it JWST}.

The {\it JWST} North Ecliptic Pole (NEP) Time-Domain Field (TDF) is a \(\sim 0.1\) sq. degree area within the {\it JWST} northern Continuous Viewing Zone \citep{jansen17}. The FLARE project intends to take deep \((\sim 27.3\) AB mag) observations with {\it JWST} 
Near Infrared Camera (NIRCam) for at least three years, utilizing the F150W ($\lambda_c \sim 1.501 \mu$), F200W (\(\lambda_c \sim 1.989 \mu\)), F356W ($ \lambda_c \sim 3.568 \mu$) and F444W (\(\lambda_c \sim 4.408 \mu\)) filters. Mapping the TDF with NIRcam will be repeated with a \(\Delta t \sim 90\) days cadence in the observer's frame in order to find (and follow-up, if possible) transients.  

The proposed series of observations would map a field-of-view (FoV) of 
$\sim 300$ arcmin$^2$ (0.083 deg$^2$) area down to at least $\sim 27.3$ AB-magnitude in all four NIRCam filters. More technical details on the proposed observations can be found in \citet{wang17}. 

In the following we use these basic observational constraints to simulate photometric data for a sample of SNe taken with the four JWST NIRCam filters listed above. We evaluate the number of potentially detectable SNe, and explore the possibilities for estimating their redshifts as well as extending the Hubble-diagram for Type Ia SNe 
toward $z \sim$~3 -- 4.
A similar study on the planned {\it WFIRST} Supernova Survey can be found in \citet{houn17}.


\section{Star formation at high redshifts}\label{sec:sfr}


For the rest of this paper we adopt the standard $\Lambda$-CDM cosmology with the following parameters:  \(\Omega_m=0.315, \Omega_\Lambda=0.685, H_0=67.4\)  \citep{planck18}, applying the {\tt astropy.cosmology} module in Python \citep{2013A&A...558A..33A}.

The first natural question that needs to be answered is the number of SNe detectable at redshifts beyond $z \sim 2$.  In order to predict this number one must know the cosmic star formation rate (SFR) at such high redshifts. 


To date numerous forms of parametrized functions have been proposed to represent the redshift dependence of the cosmic SFR (see e.g. the references given in Section~1). 
In the following we adopt and use the parametrization given by \citet{hopkins06}: 
\begin{equation}
SFR(z) = K \cdot \frac{(a+b z)h}{1+(z/c)^d}, 
\label{eq:sfr-1}
\end{equation}
where $h=H_0/100$, and $a=0.017$, $b=0.13$, $c=3.3$, $d=5.3$ are assumed following \citet{hopkins06}. Because we primarily focus on supernova rates, the $K$ factor is constrained by scaling the theoretical SN rates derived from the SFR to the observed SN rates (see below).

Eq.~\ref{eq:sfr-1} is adopted up to $z \sim 3$. Between $3 < z < 10$ we
use the observational constraints given by \citet{oesch15} based on the
UV luminosity function of high-redshift galaxies observed with the Hubble Space Telescope (HST). The redshift dependence of the SFR in this interval is  
%
%
\begin{equation}
SFR(z) \propto (1+z)^{-3.6},
\label{eq:sfr-2}
\end{equation}
which is scaled to match the Eq.~\ref{eq:sfr-1} at $z=3$. Eq.~\ref{eq:sfr-2} represents the upper limit of the observed high-$z$ SFR above $z \sim 8$ \citep{oesch15}. 
For the lower limit at $z \geq 8$ we adopted 
%
\begin{equation}
SFR(z) \propto (1+z)^{-10.4}. 
\label{eq:sfr-3}
\end{equation}

%
\begin{figure}
    \centering
    \includegraphics{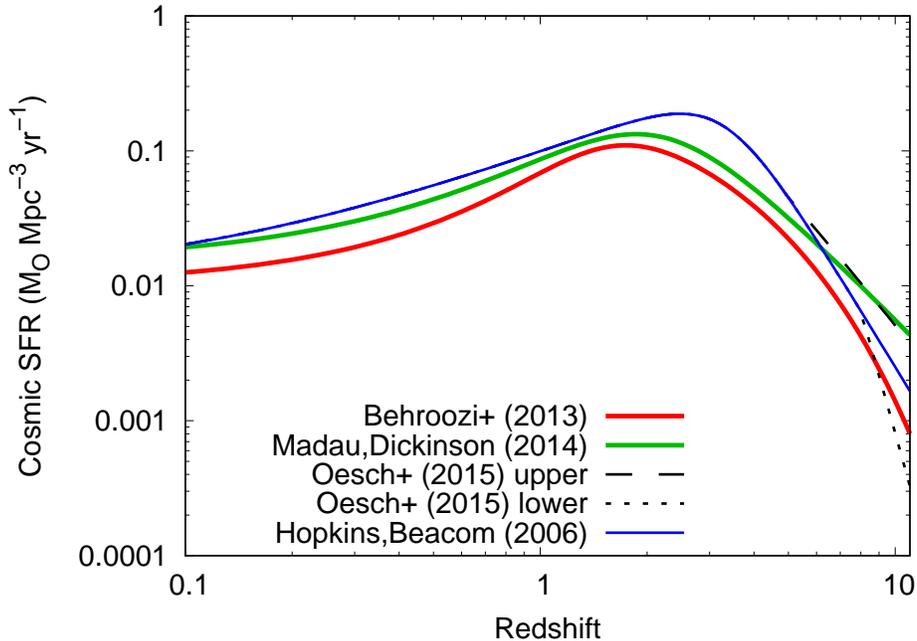}
    \caption{Parametrized forms of the cosmic SFR as a function of redshift. The \citet{hopkins06} form plus the extension based on \citet{oesch15} is adopted for further calculations, but the other forms would have given very similar results.}
    \label{fig:csfr}
\end{figure}

Several other forms of the high-redshift SFR are available over \(0 < z < 8\). For example, \citet{madi14} proposed
\begin{equation}
SFR(z) = 0.015 \cdot {\frac{(1+z)^{2.7}}{1+\left( (1+z)/{2.9}\right)^{5.6}}}  ~~(\Msol \mathrm{yr}^{-1}\mathrm{Mpc}^{-3}), 
\end{equation}
while \citet{behroozi13} obtained 
\begin{equation}
    SFR(z) = {\frac{C}{10^{A(z-z_0)} + 10^{B(z-z_0)}} },
\end{equation}
with $C = 0.18$, $A = -0.997$, $B = 0.241$ and $z_0 = 1.243$ in the unit of Eq. 4.

All of these functions give very similar redshift dependence of the cosmic SFR, as illustrated in Fig.~\ref{fig:csfr}. They predict a peak around $z \sim 2$-3 and a declining trend toward both lower and higher redshifts. 


%

\section{Supernovae at high redshifts}\label{highz}

In this section we explore the detectability of the two brightest classes of supernovae, namely SLSNe and Type Ia SNe beyond $z \sim 2$ with {\it JWST} NIRCam. 
We use model Spectral Energy Distributions (SEDs) for these SN types close to maximum light to predict the observed fluxes for redshifted SNe in the bandpasses covered by the NIRCam filters. 

\subsection{Superluminous Supernovae}\label{slsn-highz}

SLSNe are the brightest SNe known to date; they can reach or outshine $-21$ mag in any wavelength bands in the optical or near-ultraviolet \citep{quimby11, galyam12}. Observationally they can be classified into two, maybe three subclasses: members of the SLSN-I class do not show hydrogen in their spectra, unlike the hydrogen-rich SLSN-II events (note that in recent literature the hydrogen-poor SLSN-I events are often referred to simply as SLSNe, which might be a source of potential confusion, because the statistical properties of the two subclasses are systematically different, see below). 
There might be a third, very rare class, named SLSN-R, that also contains hydrogen-poor objects whose slowly evolving light curves are thought to be powered by extreme amount ($\sim 5$ M$_\odot$) of radioactive $^{56}$Ni \citep{galyam12}. Such extreme amount of $^{56}$Ni could be produced in a very massive core-collapse event induced by pair instability \citep[e.g. SN~2007bi,][]{galyam09}. 
The powering mechanism of the first two subtypes is still debated: several models including magnetar spin-down \citep[e.g.][]{kb10, nicholl17}, or interaction with hydrogen-poor circumstellar shell \citep{manos12, manos13} have been proposed, but none of them are able to fully explain all observational aspects of SLSNe.  

SLSN-I events are usually found in low-mass, metal-poor host galaxies that most often show extremely strong emission features \citep{neill11, lunnan14, leloudas15, perley16}. In this respect SLSNe-I are similar to LGRBs that also tend to prefer low metallicity hosts \citep[e.g.][]{ln06,wb06,perley16}. 
SLSNe-II, however, do not seem to show this trend: they can appear in galaxies that have broader range of mass and metallicity \citep{perley16}. 

\begin{figure}[ht]
\includegraphics[width=8.5cm,clip]{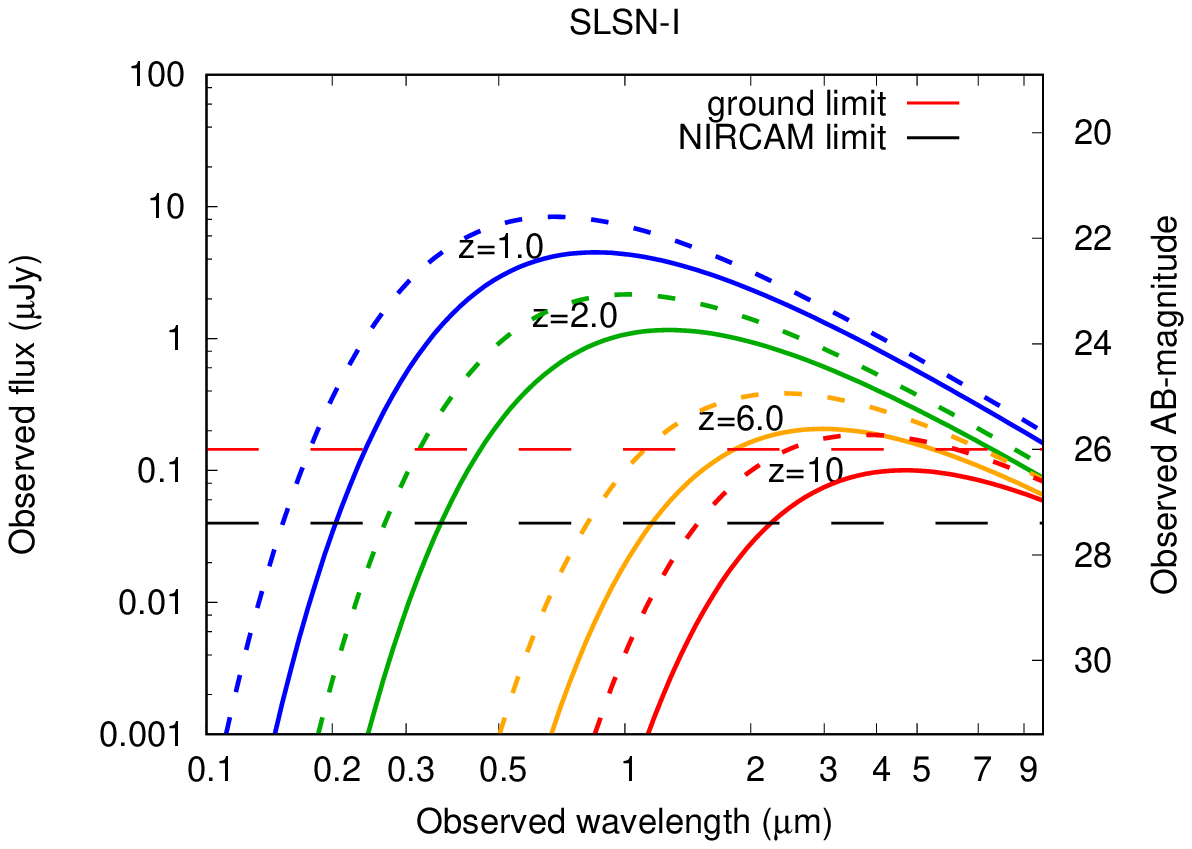}
\includegraphics[width=8.5cm,clip]{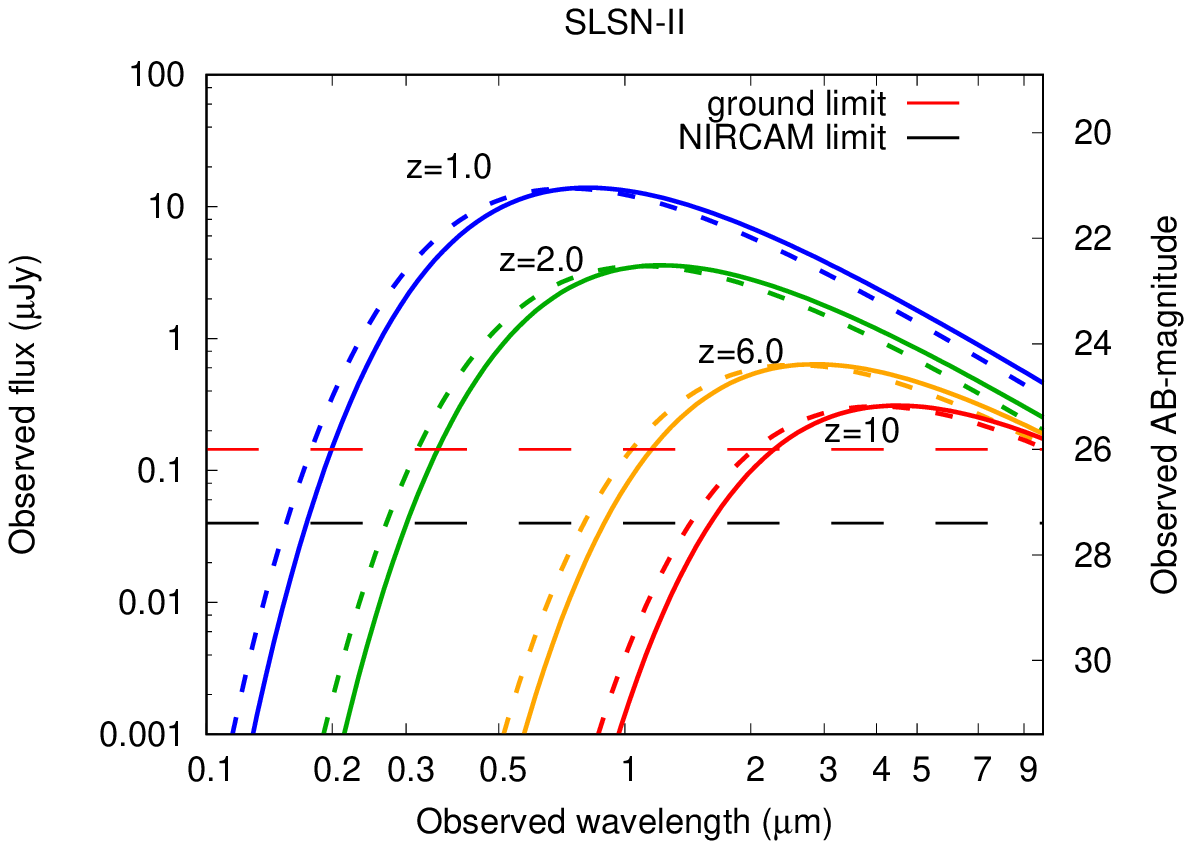}
\caption{Modeled SEDs of SLSNe-I (left panel) and II (right panel) at different redshifts. \label{fig:slsn-sed}}
\end{figure}

Figure~\ref{fig:slsn-sed} shows the blackbody-fitted SEDs of SLSNe at peak brightness shifted to various redshifts. These SEDs were constructed by combining observed, flux-calibrated spectra of various SLSNe \citep[see][for details]{wang17}. It is seen that both SLSN-I and SLSN-II events, in principle, are expected to be brighter than the NIRCam detection limit in the FLARE survey ($\sim 27.3$ AB-mag) up to $z \sim 10$, in good agreement with the results by \citet{tanaka13}. Based on this prediction, in Section~\ref{snrates} we estimate the expected number of SLSNe during the FLARE survey time.


\subsection{Thermonuclear Supernovae (Type Ia SNe)}\label{snia-highz}

%

\begin{figure}[ht]
\includegraphics[width=8.5cm]{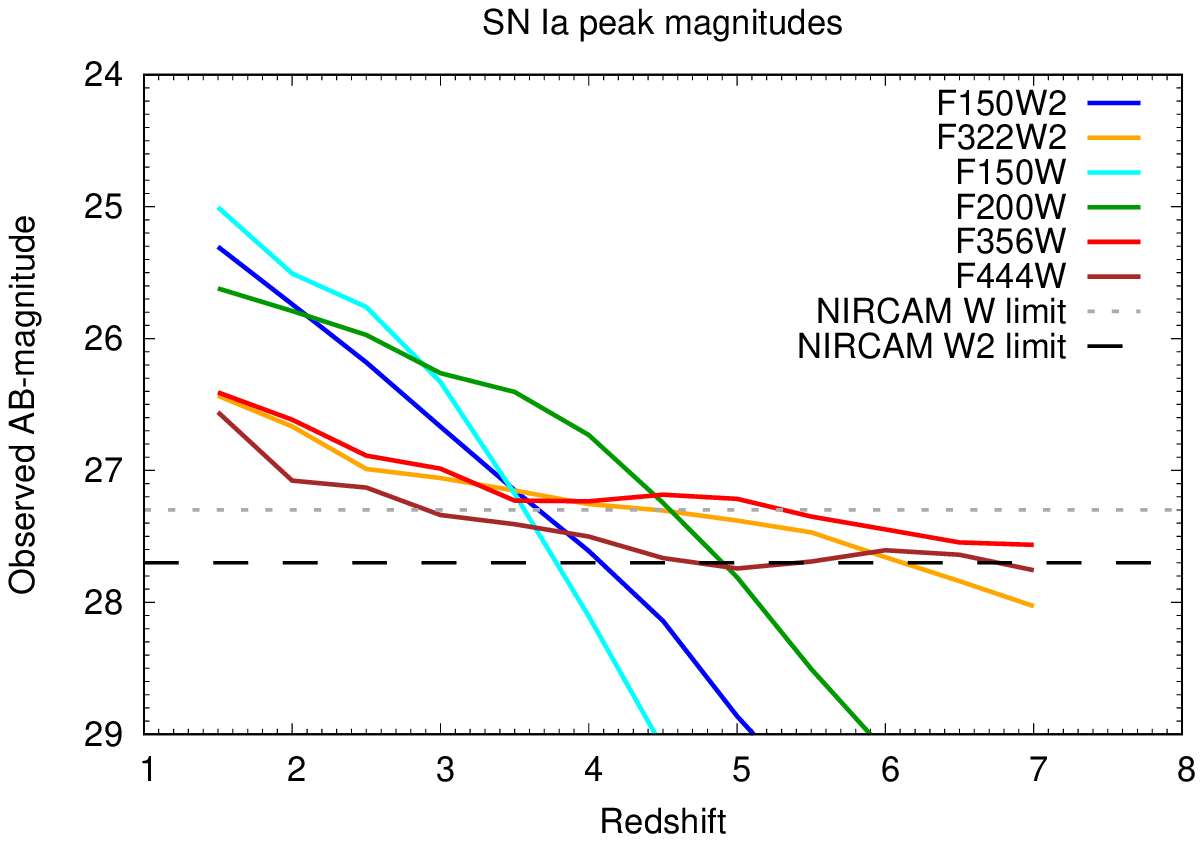}
\includegraphics[width=8.5cm,clip]{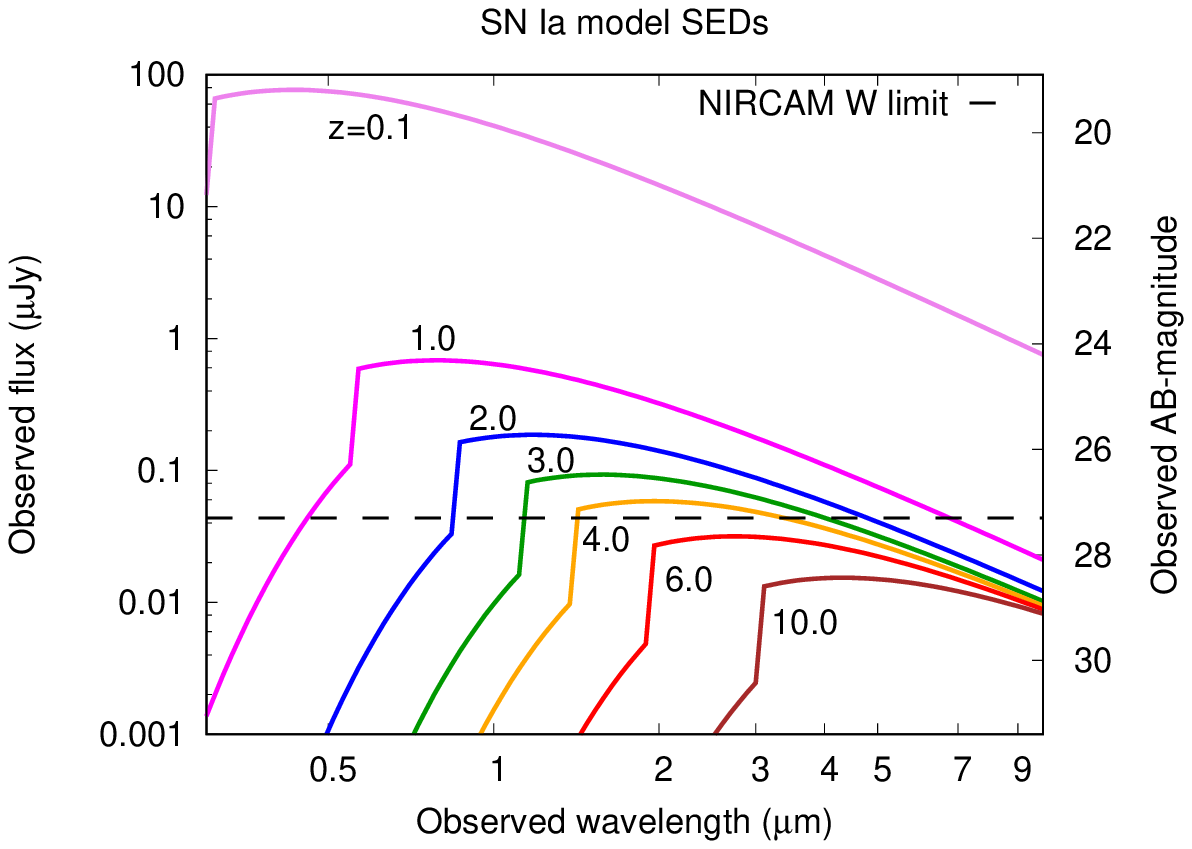}
\caption{Left panel: Peak magnitudes of Ia SNe in the NIRCAM passbands as a function of redshift. Right panel: Model SEDs of SNe Ia at different redshifts. }
\label{fig:Iased}
\end{figure}

The left panel in Figure~\ref{fig:Iased} shows the observable peak AB-magnitudes of SNe Ia with the four {\it JWST} NIRCam filters as above plus two broadband NIRCam W2 filters centered at $\sim 1.5$ and $\sim 3.2$ microns. These curves were calculated from synthetic photometry using the above mentioned {\it JWST} filter bandpasses on the Hsiao-templates for Type Ia SNe \citep{2007ApJ...663.1187H}.
It is seen that SNe Ia are expected to reach the FLARE detection limit in several NIRCam bands up to $z \sim 4$. The right panel of Fig.\ref{fig:Iased} illustrates the same conclusion by showing the blackbody-fitted SEDs of SNe Ia \citep{wang17} at various redshifts. Thus, detections of SNe Ia with {\it JWST} NIRCam is feasible in the redshift range of $1 < z < 4$. 


\section{The volumetric SN rate at high redshifts}\label{snrates}

For the cosmic star formation rate we apply the form defined by \citet{hopkins06} (Equation~\ref{eq:sfr-1}) with the extension between 
$3 < z < 8$ as in Equation~\ref{eq:sfr-2} and for $z > 8$ as in 
Equation~\ref{eq:sfr-3} \citep{oesch15} (see Section~\ref{sec:sfr} for details).

\subsection{SLSNe}

\begin{figure}
    \centering
    \includegraphics[width=8.5cm]{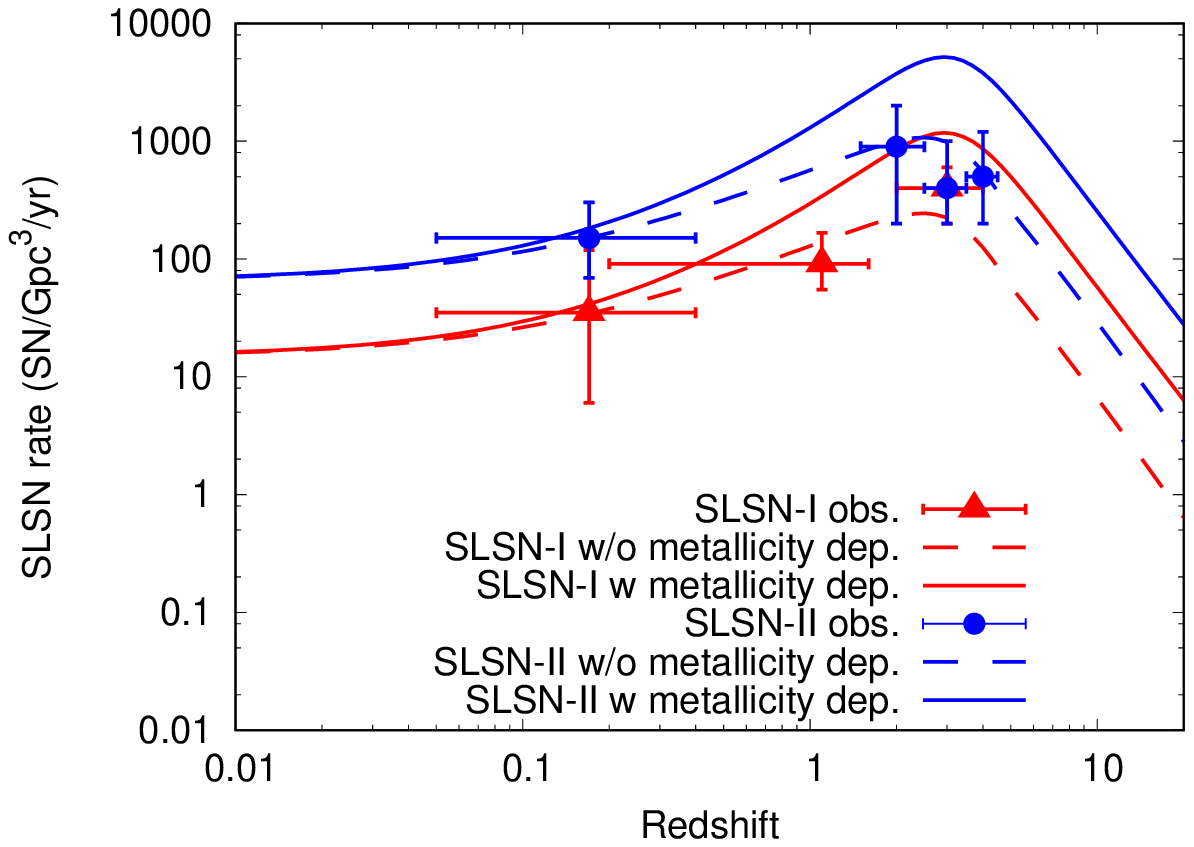}
    \includegraphics[width=8.5cm]{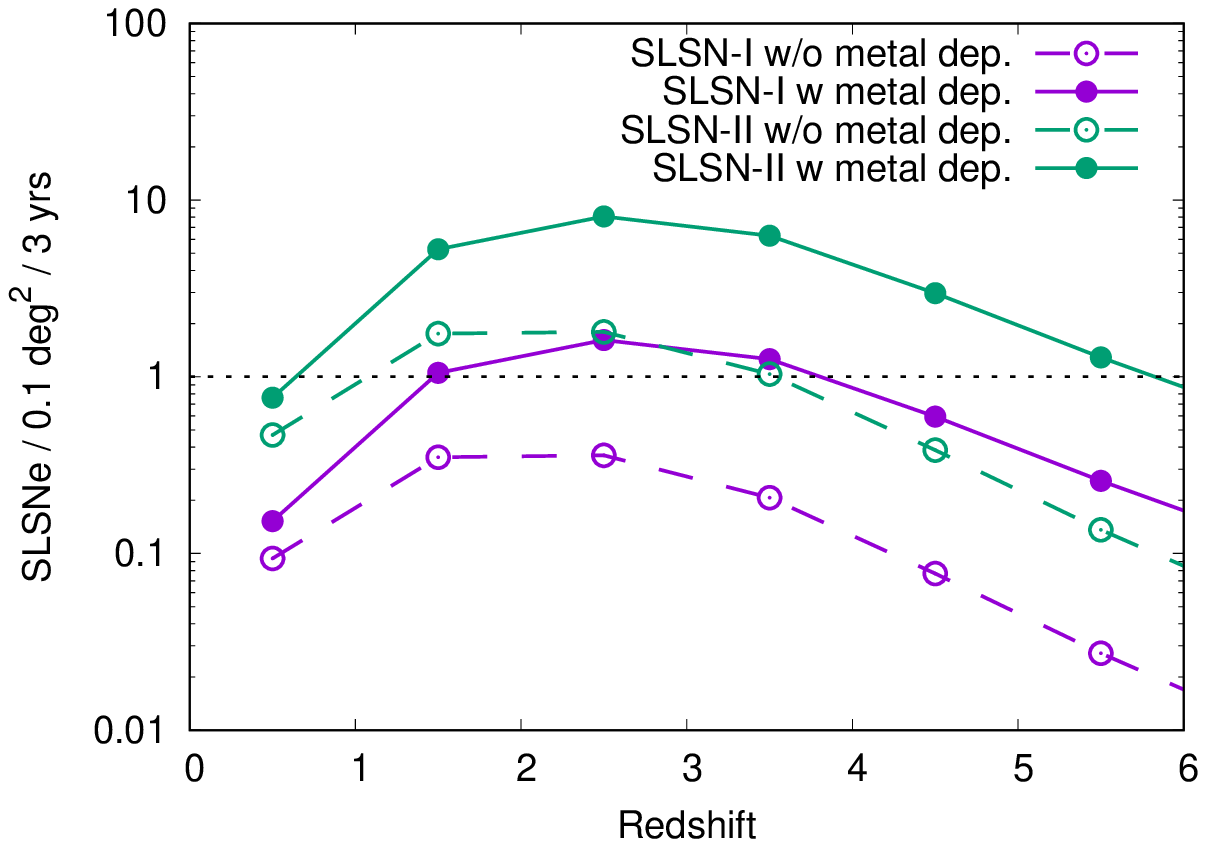}
    \caption{Left panel: Assumed volumetric rates of SLSNe-I and II as a function of redshift. Right panel: the predicted number of SLSNe in the survey field at different redshifts during the proposed 3-year-long survey.}
    \label{fig:slsne-rates}
\end{figure}

\begin{table}
    \centering
    \caption{The expected numbers of SLSN-I and II events during the survey.}
    \label{tab:slsne-numbers}
    \begin{tabular}{ccccc}
    \hline
    \hline
    Redshift & N(I)\tablenotemark{a} & N(I)\tablenotemark{b} & N(II)\tablenotemark{a} & N(II)\tablenotemark{b} \\
    \hline
    0.5 & 0 & 0 & 0 & 1 \\
    1.5 & 0 & 1 & 2 & 5 \\
    2.5 & 0 & 2 & 2 & 8 \\
    3.5 & 0 & 1 & 1 & 6 \\
    4.5 & 0 & 1 & 0 & 3 \\
    5.5 & 0 & 0 & 0 & 1 \\
    \hline
    \end{tabular}
    \tablenotetext{a}{without metallicity correction}
    \tablenotetext{b}{with metallicity correction}
\end{table}

Since SLSNe are thought to originate from very massive stars, there is practically no delay time between the formation of their progenitors and the explosion. However, the local (low-$z$) observed rates for SLSNe-I are probably biased by the fact that they tend to occur only in low-metallicity hosts, similar to LGRBs (Section~\ref{slsn-highz}). Because at high redshifts low-metallicity host galaxies are more abundant, the SLSN-I rates at $z>2$ are expected to be boosted up with respect to a rate that is estimated simply by extrapolating the local observed rate with the $SFR(z)$ function. Thus, the observed SLSN rate per redshift bin $dz$ can be expressed as
\begin{equation}\label{eq:slsn1}
\dot N_{SLSN}(z) = \frac{\dot n_{SLSN}(z)}{1+z}\,\frac{\mathrm{d}V}{\mathrm{d}z},
\end{equation}
where $\mathrm{d}V/\mathrm{d}z$ is the comoving volume,
\begin{equation}\label{eq:slsn2}
\dot n_{SLSN}(z) = \epsilon_Z(z)\,SFR(z)
\end{equation}
is the comoving rate of SLSNe, $SFR(z)$ is the cosmic star formation rate, and $\epsilon_Z(z)$ is the redshift-dependent 
efficiency factor that corrects for the metallicity dependence. 
For SNe with negligible metallicity dependence (e.g. SLSN-II), 
$\epsilon_Z(z) \approx 1$. 

The effect of metallicity on the rates of high-redshift GRBs has been extensively studied in the literature. For example, using model grids of single star progenitors of LGRBs, \citet{yoon06} computed the redshift-dependent GRB rate by using metallicity-dependent SFR and adding binaries to the collapsar model of the LGRB progenitors. From the metallicity-dependent star formation history, the observed mass function and the mass -- metallicity relation they computed the expected GRB rate as function of metallicity and redshift. More recently, many studies found that the metallicity dependence can be parametrized simply by multiplying the $SFR(z)$ function with the efficiency factor $\epsilon(z) = (1+z)^\beta$, where $\beta \approx 1.2$ \citep{kistler09, virgili11, re12, trenti13}.  

Based on the rate modelling of GRBs by \citet{trenti13}, \citet{wang17} applied the DRAGONS semi-analytic galaxy formation model \citep{mutch16} to estimate the expected number of SLSNe at high redshifts. As the progenitor models are poorly constrained, they considered simple empirically motivated models using the mean stellar metallicity of every galaxy at each simulated redshift to calculate the SLSN production efficiency factor, using stellar evolution simulations similar to \citet{yoon06} for each galaxy and average over all galaxies at that redshift. Assuming strong metallicity dependence they obtained a metallicity correction for the SLSN rate that is similar to that of the GRBs found previously (see above). In particular, they found that the peak of the SLSN rate shifts significantly toward $z \sim 5$ if strong metallicity dependence is assumed with respect to the peak at $z \lesssim 3$ when no metal dependence is used.  

Recently \citet{prajs17} calculated the volumetric rate of SLSNe at $z \sim 1$. They also estimated the rate of ultra-long GRBs based on the events discovered by the {\it Neil Gehrels Swift} satellite, and showed that it is comparable to the SLSN rate, providing further evidence of a possible connection between these two classes of events. 

As the studies mentioned above explain the observed redshift evolution of the ratio of GRB rates and SFR as $\sim (1+z)^{1.2}$ (with some increment from the power law at high redshift), we use this redshift dependence for $\epsilon_Z(z)$ in Equation~\ref{eq:slsn2}, i.e. $\epsilon_Z(z) = (1+z)^{1.2}$.
As the metallicities average out through the mass function at a given redshift, this factor provides a reasonable correction for the redshift-dependence of the frequency in low- and high-metallicity host galaxies. 

For SLSNe-I we use the observed local volumetric rates published by \citet{cooke12}, \citet{quimby13} and \citet{prajs17}. 
For SLSNe-II we adopt the observational rate as given in \citet{quimby13}, which recently turned out to be consistent with the rates estimated from three SLSNe discovered at $z \sim 2$ \citep{moriya18}. Since the metallicity dependence of SLSN-II events is less pronounced as for SLSNe-I, their rates at high redshifts might be closer to the one that can be obtained by simply extrapolating their local rates toward higher redshifts along with the $SFR(z)$ function. Nevertheless, we also apply the same redshift dependent metallicity correction as above for SLSNe-II as well in order to get an upper limit for their high-$z$ rates. 

The expected number of SLSNe between redshifts $z$ and $z + \Delta z$ can be calculated by integrating Equation~\ref{eq:slsn2} to get
\begin{equation}
N_{SLSN} = \Omega T \int_z^{z+\Delta z} \frac{\dot n_{SLSN}(z)}{1+z}\,\frac{\mathrm{d}V}{\mathrm{d}z}\,\mathrm{d}z,
\end{equation}
where $\Omega$ is the survey area and $T$ is the survey time. 
The results (the number of SNe per unit redshift interval within the survey area during the total survey time) are shown in Table~\ref{tab:slsne-numbers}, where the columns list the predicted number of SLSN-I and II events with and without the metallicity correction.


The left panel of \ref{fig:slsne-rates} displays the  adopted
SLSNe volumetric rates as a function of redshift. The right panel 
shows the predicted numbers of SLSNe in the survey field at 
different redshifts with and without the assumed redshift-dependent metallicity correction. It is seen that even if we continue the
survey up to 3 years in the observer's frame, we can expect only
very few SLSNe-I at relatively low ($z \sim 2$-3) redshifts. SLSNe-II look to be more abundant than SLSNe-I, thus, it is more probable that the newly discovered high-$z$ SLSNe will be SLSN-II events, especially if their rate also (at least slightly) depends on the host metallicity. On the other hand, even though SLSNe can be potentially detectable with {\it JWST} up to $z \sim 10$, their very low volumetric rate make
them less suitable for constraining the cosmic SFR at $z \gtrsim 5$, at least with the relatively small-area survey considered in this paper.  

Note that the uncertainties in the cosmic SFR as well as in the observed SLSN rates make all the predictions on the expected SLSN numbers uncertain by at least a factor of $\sim 2$.


\subsection{Type Ia SNe}

The volumetric rate for Type Ia SNe is different from that of core collapse events. Since the progenitors of Type Ia SNe are binaries containing at least one white dwarf (i.e. evolved) star, a significant delay time between their formation and the SN explosion is expected. Therefore, their rate can be expressed as
\begin{equation}\label{eq:iarate}
R_{Ia}(t) = \nu \int_{t_F(z_F)}^t SFR(t')\,\Psi_{DTD}(t-t')\,\mathrm{d}t' \hskip 0.5in (z_F=10)
\end{equation}
where the
efficiency \(\nu\) is the number of SNe formed per unit stellar mass (\(\Msol^{-1}\)), that is the
fraction of white dwarfs in the 3-8 \(\Msol\) range, 
$z_F$ is the formation redshift and  $ \Psi_{DTD} $ is their delay time distribution.

%
%
%
%
%


For the delay time distribution various models are proposed in the the theoretical and observational literature.
%

In the SD scenario (corresponding to short delay times), from
simple analytic modelling of main-sequence lifetime as a function of mass \citep[e.g.][]{barbaryphd} one can get 
\begin{equation}
\Psi(t) \propto t^{-0.46},
\end{equation} 

On the other hand, double degenerates (long delay) result in 
\begin{equation}
\Psi(t) \propto t^{-1}.
\end{equation}

Population synthesis models predict universal DTD shapes for SNe Ia,
independent of the details of common envelope prescription, mass transfer rate, hydrogen retention efficiency or metallicities \citep[]{nelemans13, moe13}. 
For example, \citet{strolger04} considered two general forms for the DTD to explain the redshift distribution of SNe Ia discovered in the {\it Hubble} Higher-z Supernova Search program between $0.2 < z < 1.6$. They applied exponential distributions like 
\begin{equation}
\Psi(t) = e^{-t/\tau}/\tau
\end{equation}
or Gaussian distributions 
\begin{equation}
\Psi(t) = e^{-(t-\tau)^2/{2\sigma^2}}/{\sqrt {2\pi}\sigma
}
\end{equation}
assuming either wide ($\sigma = 0.5 \tau$) or narrow ($\sigma = 0.2 \tau$) Full Width at Half Maximum (FWHM) for the latter. Also, the peak of the Gaussian distributions were set in between $0.2 < \tau < 10$ Gyr. \citet{strolger04} found $\tau \sim 4$ Gyr as their best-fit value. 

Although at present most of the observational evidence point toward the $\Psi(t) \sim t^{-1}$ DTD \citep[e.g.][]{maoz14}, in this paper we consider all the DTD forms mentioned above to predict the expected redshift distribution of SNe Ia at $z>1$ redshifts. For the delay-time parameter we assume $\tau$ = 0.5, 1.0, 2.0, 3.0 and 4.0 Gyr.

\begin{figure}
    \centering
    \includegraphics[width=0.45\textwidth]{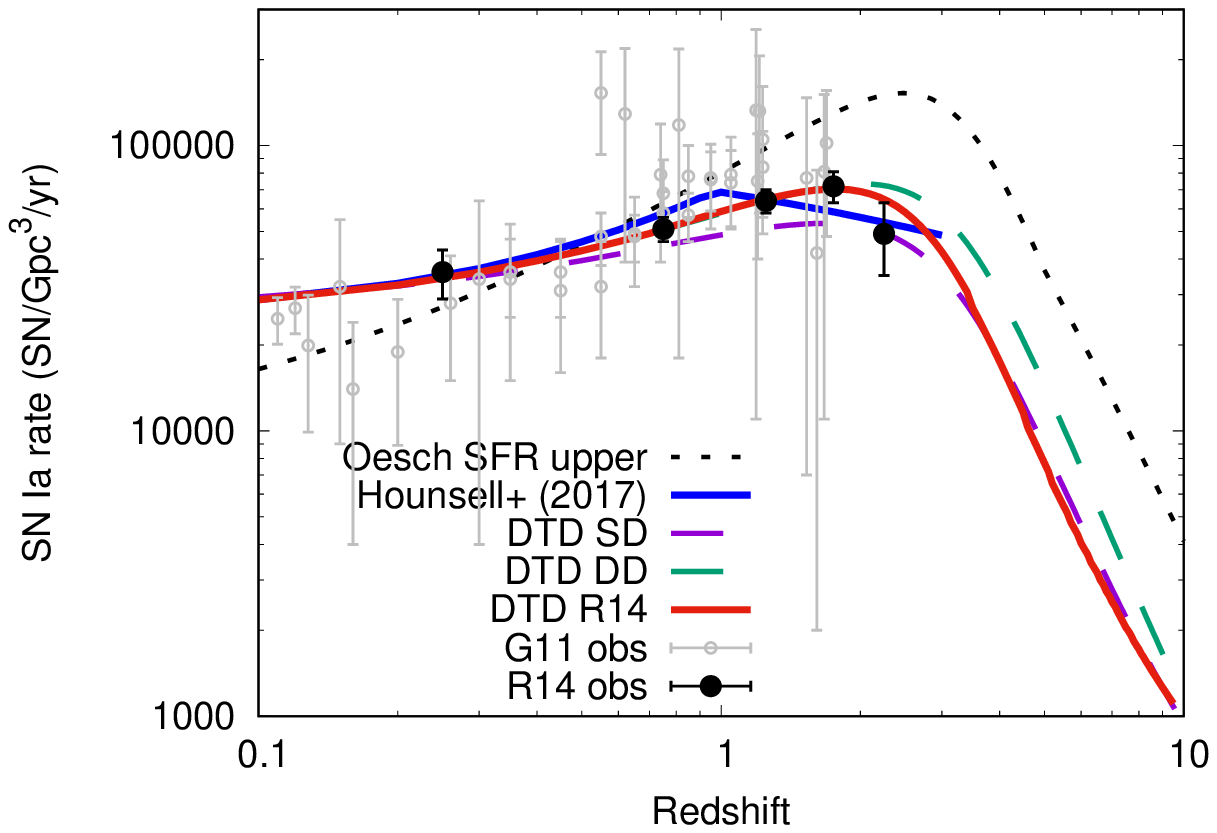}
    \includegraphics[width=0.45\textwidth]{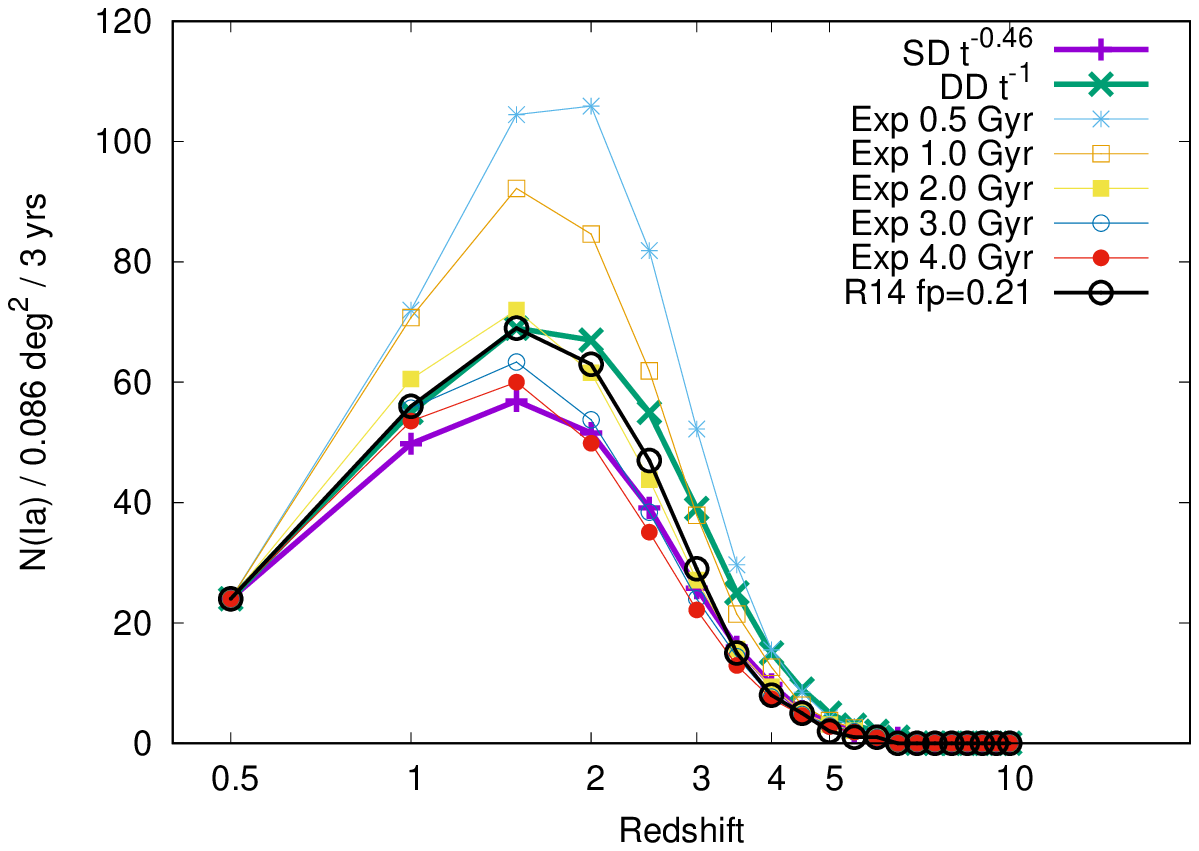}
    \includegraphics[width=0.45\textwidth]{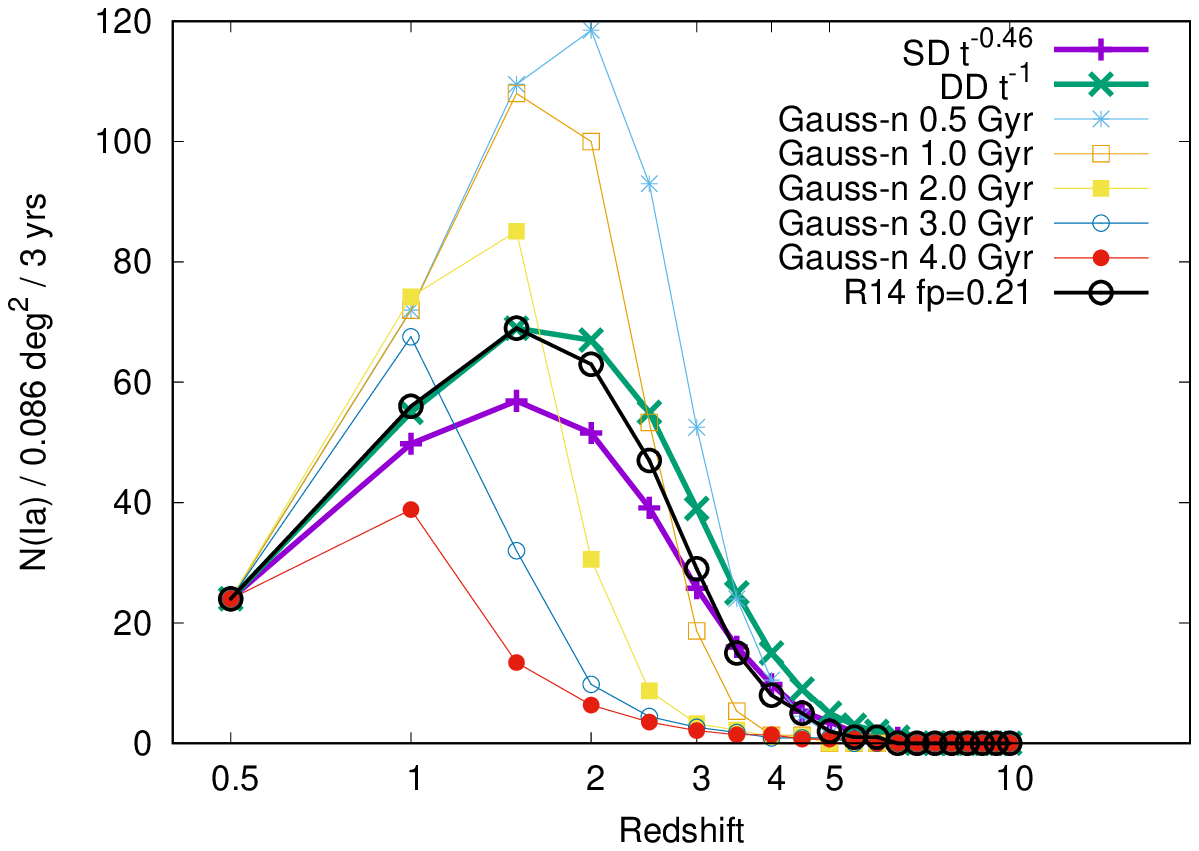}
    \includegraphics[width=0.45\textwidth]{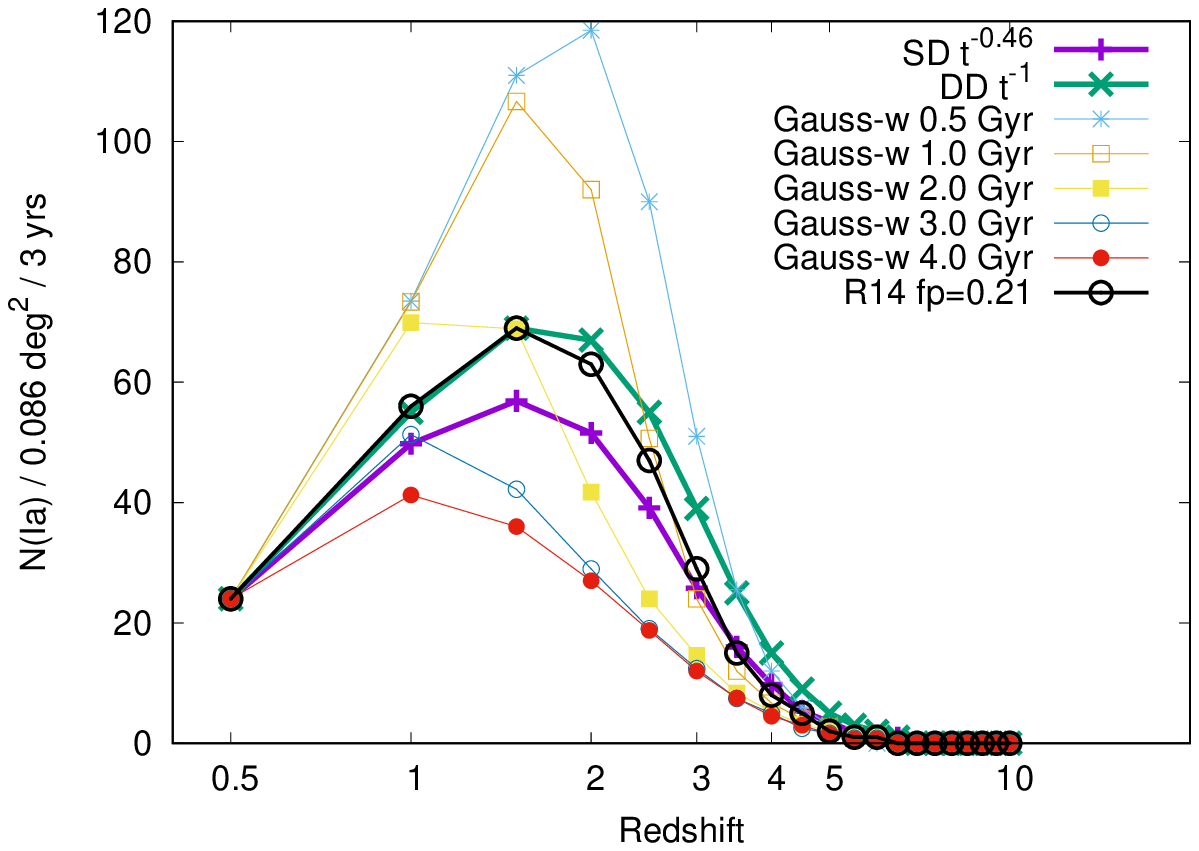}
    \caption{Top left panel: The observed SN Ia rates as a function of redshift (open circles) compared to the rates assumed in \citet{houn17} (blue curve), \citet{rodney14} (red curve) 
    and the ones calculated with different forms of DTD (see text). The assumed SFR \citep{oesch15} corresponding to negligible DTD is also shown as a dotted curve. Top right panel: the expected number of SNe during the survey time (3 years) assuming the SD, DD and exponential form of DTD (colored lines and symbols), as indicated in the legend. The black line and symbols correspond to the best-fit rates given by R14. All curves are normalized to the same value at $z=0.5$. Bottom left panel: the same as the top right panel, but with the Gauss-narrow DTD. Bottom right: the same as the top right panel, but with the Gauss-wide DTD.}
    \label{fig:rates-comp}
\end{figure}

In Figure~\ref{fig:rates-comp} the top left panel shows the observed SN Ia rates collected from literature: the grey symbols are from mostly ground-based observations \citep[][and references therein]{graur11}, while the black filled circles are the final binned rates from the CLASH and CANDELS surveys \citep[][R14 hereafter]{rodney14}. The continuous lines represents various theoretical rates corresponding to different DTD forms listed above. Thick red line shows the best-fit SN Ia rate given by R14, which is basically a $t^{-1}$ DTD scenario with a fraction of ``prompt'' Ia population, $f_p$, mixed in:
\begin{equation}
\Psi(t)~=~
\begin{cases}
0 & (t<0.04~\mathrm{Gyr}) \\ 
7.132 \cdot \eta  \cdot f_p (1-f_p)^{-1} & (0.04 < t < 0.5~\mathrm{Gyr}) \\
\eta \cdot t^{-1} & (t>0.5~\mathrm{Gyr}), 
\end{cases}
\label{eq:R14}
\end{equation}
where $\eta = 2.25$ and $f_p = 0.21$ were adopted as the best-fit parameters from R14.
The thick blue line denotes the parametrized SN Ia rate applied by \citet{houn17} (H17) for estimating the number of SNe Ia in the {\it WFIRST} Supernova Survey:
\begin{equation}
R_{Ia}(z)~=~
\begin{cases}
2.5 \times 10^{-5} (1+z)^{1.5} & (z<1) \\
9.7 \times 10^{-5} (1+z)^{-0.5} & (1 < z < 3) \\
\end{cases}
\label{eq:wfirst}
\end{equation}

It is seen that the ground-based rates are not constraining beyond $z > 1$, while the R14 and H17 models are in good agreement with the binned $HST$ rates. However, the predicted rates at $z >1$ redshifts are uncertain. The new SN Ia discoveries beyond $z > 2$ would provide critical and unprecedented information on the real nature of the DTD, for example, a better estimate for the fraction of the ``prompt'' Ia population.   

The other panels in Figure~\ref{fig:rates-comp} plot the expected numbers of SNe Ia in the $\sim 300$ arcmin$^2$ survey field during the total survey time (3 years) assuming the various DTD functions detailed above.
All numbers are normalized to the same value in the first redshift bin centered at $z = 0.5$. Black circles show the R14 rates with ``prompt'' fraction of $f_p = 0.21$, which we use as the best-fit reference rates at $z > 1$. It is seen that the various DTDs predict roughly similar redshift dependence at $z>1$, although the predicted number of SNe Ia may differ by a factor of $\gtrsim 2$ around $z \sim 2$.   Table~\ref{tab:snia-numbers} lists these numbers for three cases: the SD, DD and the R14 scenarios, as shown above.

The uncertainties of the predicted numbers of SNe Ia are difficult to estimate because of the high uncertainty of the ground-based data and the low number of observational constraints above $z \sim 1$ (Fig.~\ref{fig:rates-comp}). If we adopt the predictions from the SD and DD scenarios ``as is'', then their difference may be used as a proxy for the uncertainty of the predicted rate at various redshift:
$\sigma_N \approx 0.5 \times |N_{SD} - N_{DD}| $.
These numbers are given in the last column of Table~\ref{tab:snia-numbers}.


\begin{table}
\centering
\caption{The expected numbers of SNe Ia in the survey field.}\label{tab:snia-numbers}
\begin{tabular}{ccccc}
\hline
\hline
Redshift & N(Ia) & N(Ia) & N(Ia) & $\sigma_N$ \\
  &(SD) & (DD) & (R14) & (SD-DD)\\
\hline  
0.5 & 24 & 24 & 24 & 1\\
1.0 & 50 & 55 & 56 & 2\\
1.5 & 57 & 69 & 69 & 6\\
2.0 & 52 & 67 & 63 & 7\\
2.5 & 39 & 55 & 47 & 8\\
3.0 & 26 & 39 & 29 & 6\\
3.5 & 16 & 25 & 15 & 4\\
4.0 & 10 & 15 & 8 & 2\\
4.5 & 5 & 9 & 5 & 2\\
5.0 & 4 & 5 & 2 & 1\\
5.5 & 2 & 3 & 1 & 1\\
6.0 & 1 & 2 & 1 & 1\\
6.5 & 1 & 1 & 0 & 1\\
7.0 & 0 & 0 & 0 & 1\\
\hline
\end{tabular}
\end{table}



\section{SN Ia simulations}
\label{sec:simul}

In this Section we aim at extending the Hubble-diagram for Type Ia SNe beyond redshift $\sim 2$. In order to discover, classify and analyze a statistically significant sample of Type Ia SNe with {\it JWST}, a robust methodology for all of these tasks is needed. In this Section we use a simulated sample of Ia SNe computed by applying the
{\tt sncosmo}\footnote{http://sncosmo.readthedocs.io/en/v1.6.x/}
code \citep{2016ascl.soft11017B} that is designed to simulate
SEDs and light curves of Type Ia SNe at any redshift. 

We simulate a 3 year-long photometric survey of SNe Ia by 
generating a sample of 320 SNe distributed in the $0<z<5$ redshift interval according to the SN~Ia volumetric rate by R14, as discussed in Section~\ref{snrates} and given in the fourth column of Table~\ref{tab:snia-numbers}.


The epoch of maximum light for each SN is distributed uniformly within the 3 year-long survey window (in the observer's frame), and observational epochs with a regular cadence of 90 days are set during the survey time. This resulted in a maximum of 12 observational epochs in this simulation.

Luminosity distances are assigned to the simulated SNe
via the {\tt astropy.cosmology} module by adopting the 
\citet{planck18} cosmology, as above. To model the SED 
of the simulated SNe as a function of time we use the 
SALT2 templates \citep{betoule14} extended to 2.5 microns \citep{houn17}, as built-in {\tt sncosmo}. These templates allow the computation
of synthetic photometry in all 4 JWST NIRcam bands up to $z \sim 5$ redshifts.


The distribution of the peak absolute brightnesses 
of the simulated SNe are approximated by assuming that
their rest-frame V-band magnitudes have Gaussian distribution
around the mean value of $M_V = -19.3$ mag and a FWHM of
$\sim 0.5$ mag \citep{2014AJ....147..118R}. Such a distribution may predict a few SNe Ia brighter than $M_V \sim -20$ mag at peak, which are not frequently observed, but could be associated with the brightest 91T/Super-Chandra Ia events \citep[e.g. 2007if,][]{yuan10}.

Beside the peak absolute magnitude, the SALT2 model also needs the stretch ($x_1$) and color ($c$) parameters to be set. The distribution of these two parameters are adopted from \citet{scolnic16}: they derived two-sided Gaussian distributions for both $x_1$ and $c$ by fitting data from large SN~Ia surveys. Here we use their fits to all data, resulting in $\langle x_1 \rangle = 0.938$, $\sigma^-(x_1) = 1.551$, $\sigma^+(x_1) = 0.269$, $\langle c \rangle = -0.062$, $\sigma^-(c) = 0.032$, $\sigma^+(c) = 0.113$ for their mean and asymmetric full-width-at-half-maximum values, respectively. 

Having the redshift ($z$), luminosity distance ($D_L$), 
moment of maximum light ($t_{max}$), rest-frame V-band
absolute magnitude ($M_V$(max)), stretch ($x_1$) and color ($c$) for each simulated SN, we compute synthetic photometry in the {\it JWST} NIRCam
F150W, F200W, F356W and F444W filter bandpasses at each
observational epoch by taking into account time dilation
and flux density corrections due to redshift. The effect of dust extinction within the host galaxy is incorporated in the model distribution of the $c$ parameter, 
while dust extinction in the Milky Way should be negligible
in these {\it JWST} bandpasses. 


\subsection{Statistics}
\label{sec:stat}

\begin{figure}[ht]
\label{fig:sim-stat}
\includegraphics[width=8.5cm]{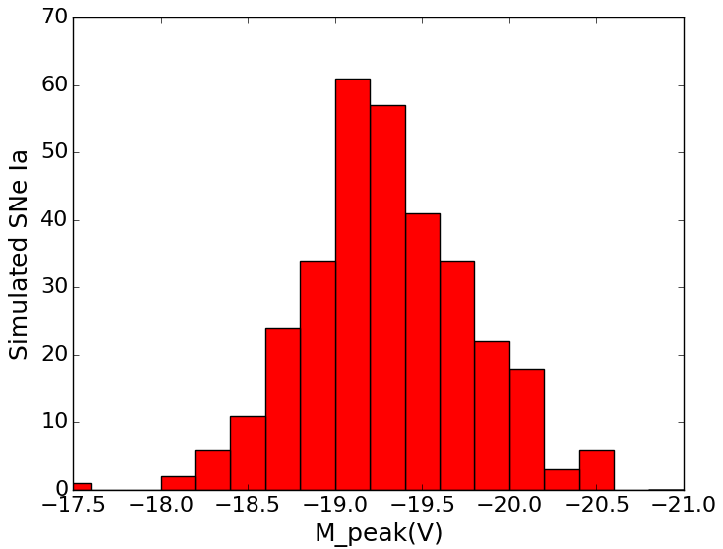}
\includegraphics[width=8.5cm]{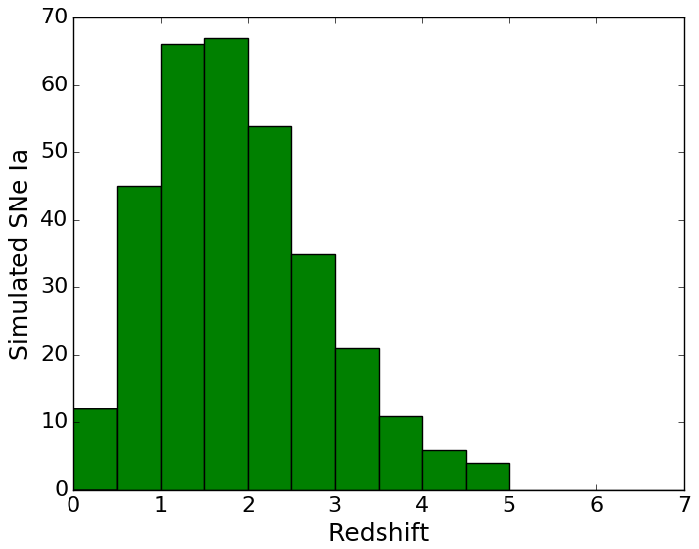}
\caption{Left panel: the histogram of the rest-frame 
V-band absolute magnitudes of the simulated SNe.
Right panel: the redshift distribution of the simulated SNe.}
\end{figure}

In the left panel of Figure~\ref{fig:sim-stat} the histogram of the V-band absolute peak brightnesses for the simulated SNe is plotted, while the right panel shows the distribution of the same SN sample in redshift space.
The distribution of peak brightnesses introduces a large scatter in the observed peak magnitudes on the Hubble diagram.
This scatter can be reduced by applying the  stretch -- or decline-rate  -- correction which is commonly
applied for SNe Ia when light curves in the rest-frame optical bands are available. However, in the proposed FLARE survey, as shown below, well-sampled light curves cannot be expected. Thus, alternative methods for taking into account and correcting for the peak brightness distribution are needed. 

We define two types of detection in our simulation. ``Strong detection'' means that a particular SN is detected (i.e. brighter than 27.3 AB-magnitude, see Section~\ref{sec:survey}) in all 4 NIRCam filters simultaneously at a given epoch. ``Weak detection'' is defined as a detection only in at least one NIRCam bandpass at a given epoch. In the full sample containing 320 simulated SNe, 48 pass the ``strong detection'' criterion on at least 1 epoch during the survey, but only 9 of them are detected on 2 epochs. 
Since low number statistics may influence the properties of the ``strong'' sample, we computed 10 different simulations having the same number of SNe whose parameters are randomly distributed within the allowed parameter range. The maximum redshift of these SNe turned out to be $z_{max} \sim 3.64 \pm 0.33$, while the average redshift of the ``strong detection'' subsample is $\langle z \rangle = 1.822 \pm 0.028$.  The number of SNe in the ``weak detection'' group is 314, 151 of which are detected on 2 epochs. The maximum redshift in the ``weakly detected'' sample was 4.98, while the mean redshift of this subsample is $\langle z \rangle = 1.921 \pm 0.041$. 

These numbers suggest that even though a significant number ($\lesssim 50$) of SN Ia detections in all 4 JWST NIRCam filter bands is expected during the proposed 3 year-long survey, only 
$\lesssim 10$ \% of them would be detected on 2 consecutive epochs. Such a sparsely sampled ``light curve'' is clearly not capable of providing the necessary correction for the peak brightness distribution via the usual stretch/decline rate measurement. Moreover, since conventional spectroscopic observations are not feasible for SNe at $z \gtrsim 2$, the determination of the redshifts of the detected SNe must rely solely on photometry. 

In the following sections we explore the possibilities and the feasibility of estimating the redshift and the true peak brightness of SNe from JWST NIRCam photometry. 

\subsection{Estimating photometric redshifts}
\label{sec:photoz}

\begin{figure}[ht]
\begin{center}
\includegraphics[]{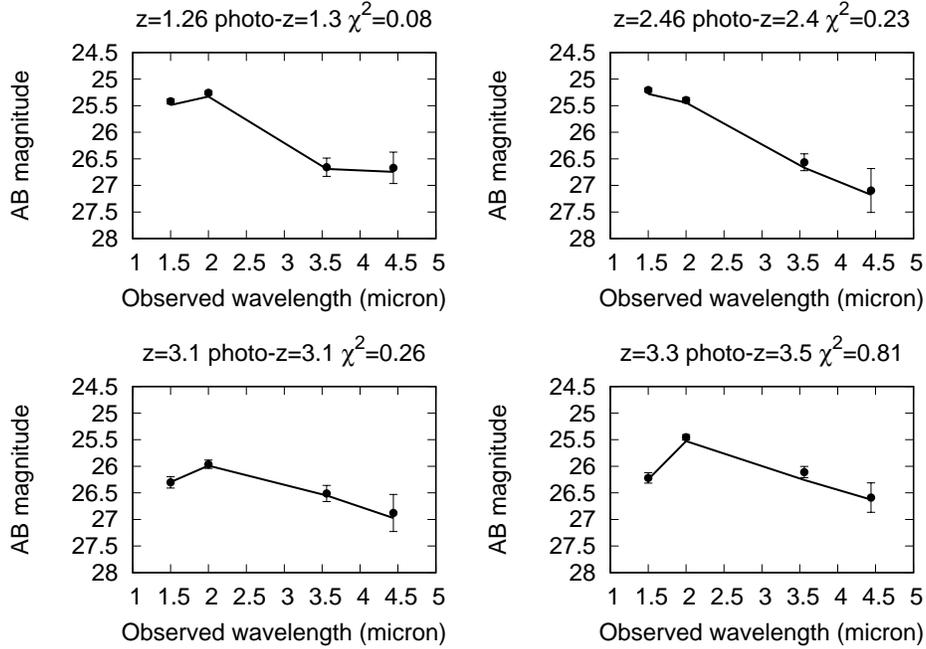}
\caption{Examples of ``observed'' SEDs (filled symbols) and synthetic magnitudes derived from the best-fit SALT2 models (lines).}
\label{fig:photoz-sed}
\end{center}
\end{figure}

\begin{figure}[ht]
\includegraphics[width=8.5cm]{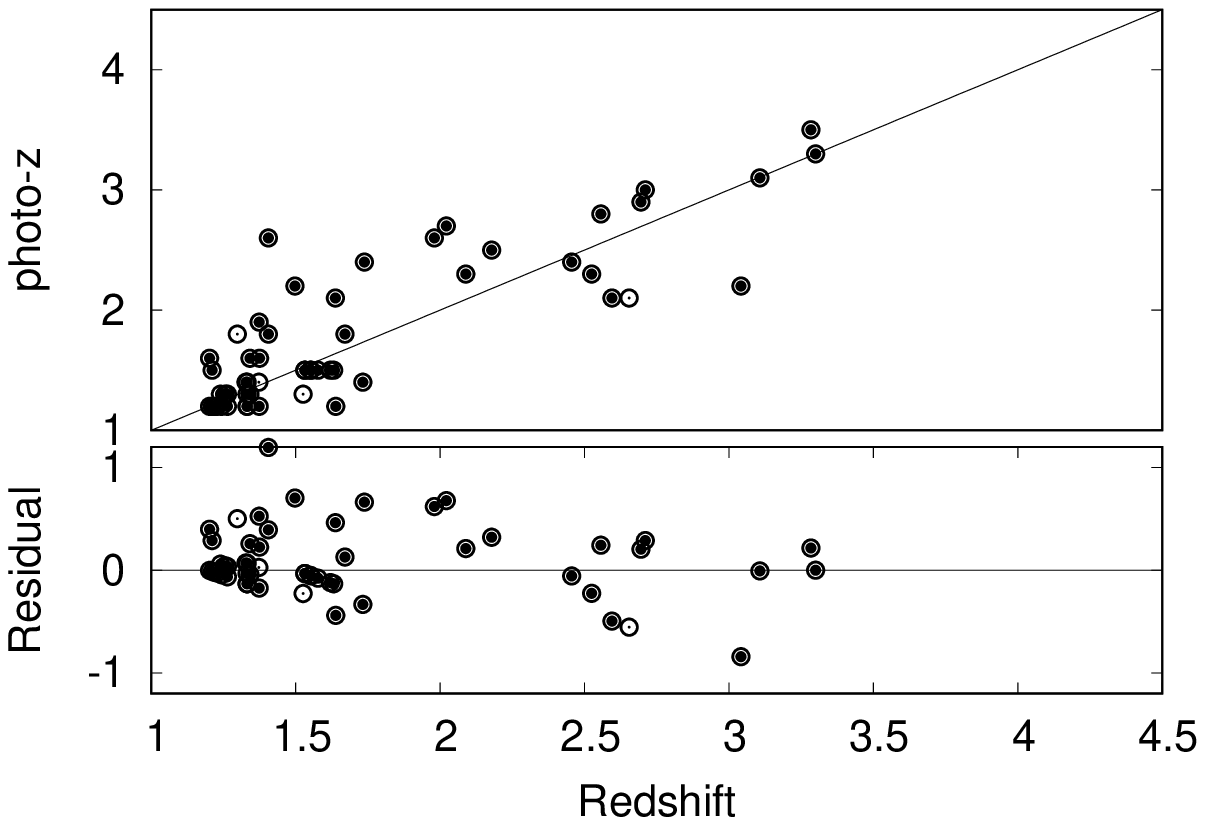}
\includegraphics[width=8.5cm]{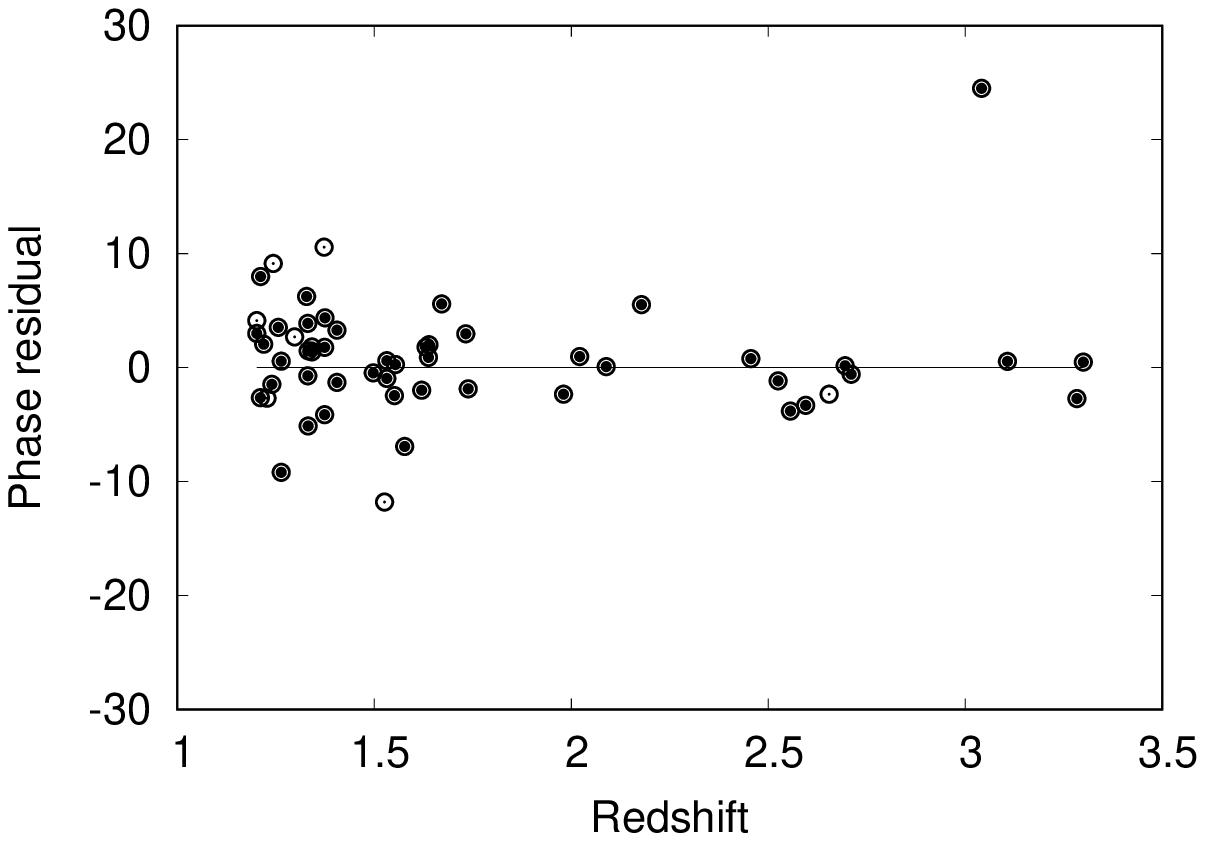}
\caption{Left panel: photometric redshifts of simulated SNe Ia derived by fitting SALT2 templates to fluxes in 4 NIRCAM filter bandpasses. Open circles denote fits having $\chi^2 < 10$, while
filled circles correspond to $\chi^2 \leq 1$. Right panel: the residual of the SN phases recovered from the SALT2 template fits. }
\label{fig:photoz-comp}
\end{figure}

In many previous studies the photometric classification of SNe Ia discovered at $z>1$ were performed via light curve fitting \citep[e.g.][]{jones13, graur14, rodney14, rodney15, rubin18}. For such light curve simulations usually the SALT2 code \citep{2007A&A...466...11G,guy10} is applied. This kind of classification, although being robust, requires not only detections in various wavelength bands, but also multi-epoch ($\gtrsim 4$) observations. Also, the reliability of this method is significantly improved if the redshift of the transient can be estimated independently, e.g. by ground-based deep spectroscopy of its host galaxy.  

As shown in the previous section, the redshifts of SNe detected with {\it JWST} in the FLARE survey must be determined from photometric/SED data.
Having accurate and precise photometric redshifts may enable the use of SNe Ia, measured only with photometry, to probe cosmology. This can dramatically increase the science return of future supernova surveys.

For example, the Large Synoptic Survey Telescope will use improved versions of the analytic photo-z estimator of \citet{wang07}
and \citet{wangetal07}. That method uses colors as well as peak magnitudes, or colors only, to estimate the redshift of SNe Ia. 
It is an empirical, model independent method (no templates used). 

Photometric redshifts derived from multi-band photometry are also proposed for thousands of SNe Ia expected from the Dark Energy Survey \citep{bernstein12}, although they preferred the photo-z estimates for the host galaxies rather than the SNe, because the co-added frames of galaxies can be $\sim 2$ mag deeper than individual SN frames. \citet{sanchez14} presented an in-depth comparison of various photo-z methods and codes available for galaxies, and estimated a $\sim 0.08$ uncertainty in photo-z for a sample of $\sim 15,000$ galaxies. 

We propose the photo-z determination for SNe Ia detected with {\it JWST} NIRCam by fitting the 4-band SEDs with the extended SALT2 templates. The fitting parameters were the epoch of maximum light ($t_{max}$), the rest-frame peak V-band absolute magnitude ($M_V$), the redshift ($z$) and
the SALT2 color parameter ($c$), while the SALT2 stretch parameter ($x_1$) was kept fixed at $x_1 = 0.938$ (see Section~\ref{sec:simul}). 
This method works for SNe Ia between $1 < z < 4$ redshifts, and some examples for the fits to the simulated SN sample are shown in Figure~\ref{fig:photoz-sed}. 
Here the flux uncertainties are derived from the wavelength-dependent flux sensitivity limits for {\it JWST} NIRCam as shown on the {\it JWST} 
website\footnote{\tt https://jwst-docs.stsci.edu/display/JTI/NIRCam+Sensitivity}. The best-fit SALT2 templates are found by simple $\chi^2$ minimization, which seems to be an adequate solution as long as the SALT2 templates can indeed model the evolution of rest-frame SEDs of high-redshift SNe Ia in a similar way as their low-redshift counterparts.
Note that since the observed NIR fluxes of SNe Ia above $z \sim 2$ are increasingly dominated by their rest-frame optical SEDs, the uncertainties in the rest-frame NIR SEDs do not bias strongly the photo-z determination, especially for $z > 3$ events when the peak of the SED shifts into the NIRCam bands (see the lower panels in  Figure~\ref{fig:photoz-sed}).

Figure~\ref{fig:photoz-comp} (left panel) compares the photo-z estimates with the ``true'' redshifts for the simulated SN sample. It is seen that the photo-z estimates are the best between the $2 < z < 3.5$ redshift interval, as explained above. For $z < 2$ most of the photo-z values are in reasonable agreement with the true redshifts, although there are some deviating SNe having $\Delta z > 1$ residuals. The overall uncertainty, estimated as the standard deviation of the $\Delta z < 0.5$ residuals (after removing the outliers that represent $\sim 20$ \% of the sample), is $\sigma_z \sim 0.25$.

It is emphasized that this accuracy can be reached only when the SN is successfully detected in all 4 NIRCam bands. Non-detection in any of these bands can degrade the quality of the fitting, thus, the accuracy of the photo-z estimate.

The right panel of Figure~\ref{fig:photoz-comp} plots the residuals between the simulated and recovered rest-frame phases (i.e. rest-frame days from epoch of maximum light) against redshift. The phases of all the simulated observations could be recovered within $\pm 10$ days. Again, the phase determination seems to work better for 
$z > 2$ SNe. The uncertainty of the phase determination, estimated as above, is found to be $\pm 4.9$ days.

In reality, core-collapse SNe are expected to contaminate the low redshift ($z <1$) sample. \citet{wang17} discussed the detection possibility of both Type II-P and Type Ib/c SNe based on the Nugent-templates \citep{nugent02}. They found that at low redshifts SNe Ib/c are too faint to be detected in the reddest NIRCam filter (F444W) in the FLARE survey (see Figure~7 and Table~2 in \citet{wang17}). Also, SNe Ib/c have lower rates than Type II SNe, which also reduces the probability of their detection during the survey time. Thus, SNe Ib/c are not likely to pass our ``strong'' detection criterion. 
Type II-P SNe, on the other hand, may appear in the survey field-of-view with a factor of $\sim 3$ lower number in the F444W filter than SNe Ia \citep{wang17}, but their NIRCam colors and color evolution looks different from those of SNe Ia (see Section~\ref{sec:color}).

It is concluded that in the $1 < z < 3.5$ redshift interval accurate flux measurements of SNe Ia with 4 {\it JWST} NIRCam filter bands allow redshift and phase estimates with $\sim 0.25$ and $\pm 4.9$ day uncertainties, respectively. Note, however, that several other disturbing circumstances are ignored in this study, like e.g. the host galaxy contamination in the SN fluxes, or non-standard extinction in the host galaxy that cannot be captured by the SALT2 color parameter. Thus, our results are somewhat optimistic, but may serve as a guideline for the real observational studies with $JWST$. 


\subsection{Color-color diagrams}
\label{sec:color}

\begin{figure}[ht]
\includegraphics[width=8.5cm,clip]{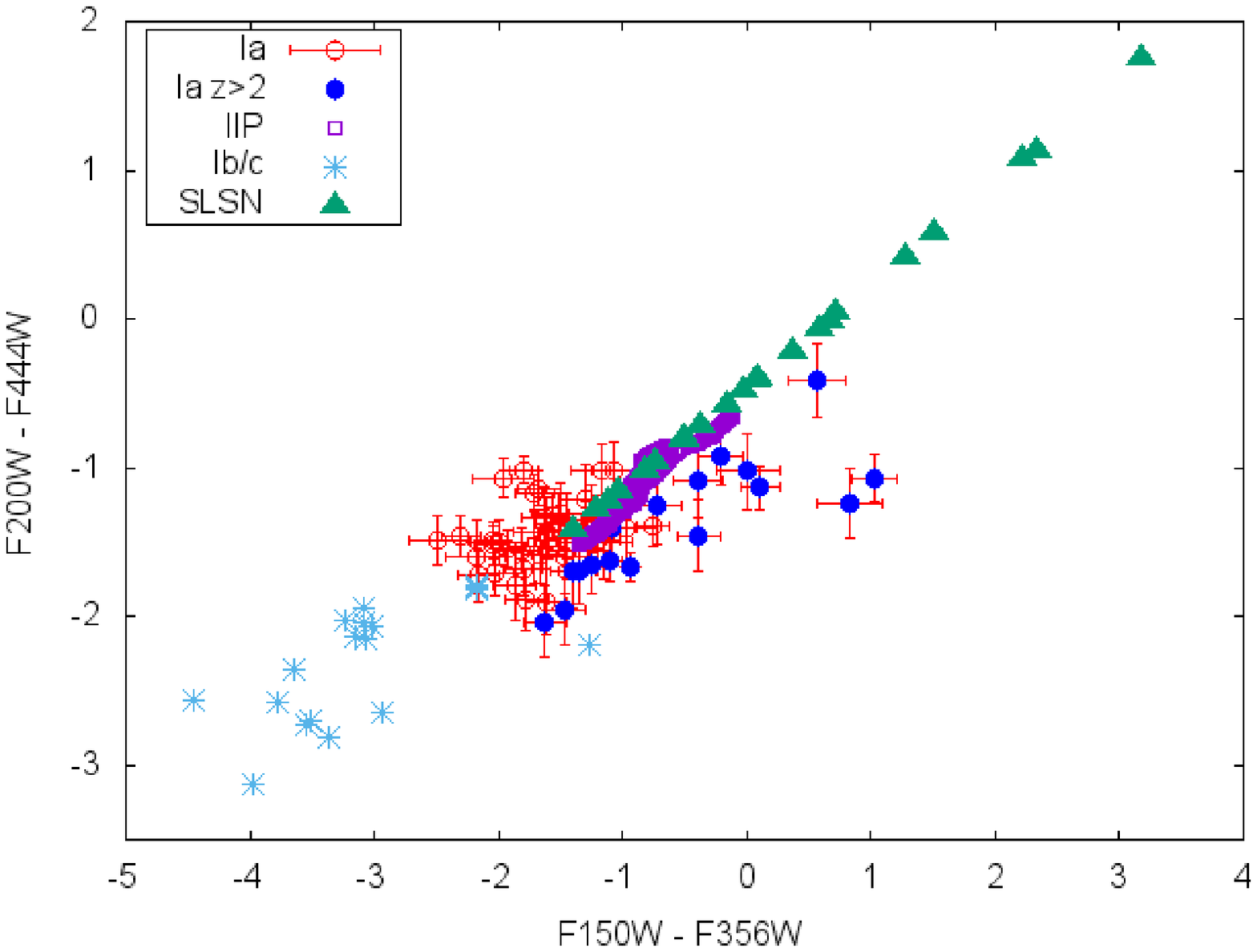}
\includegraphics[width=8.5cm,clip]{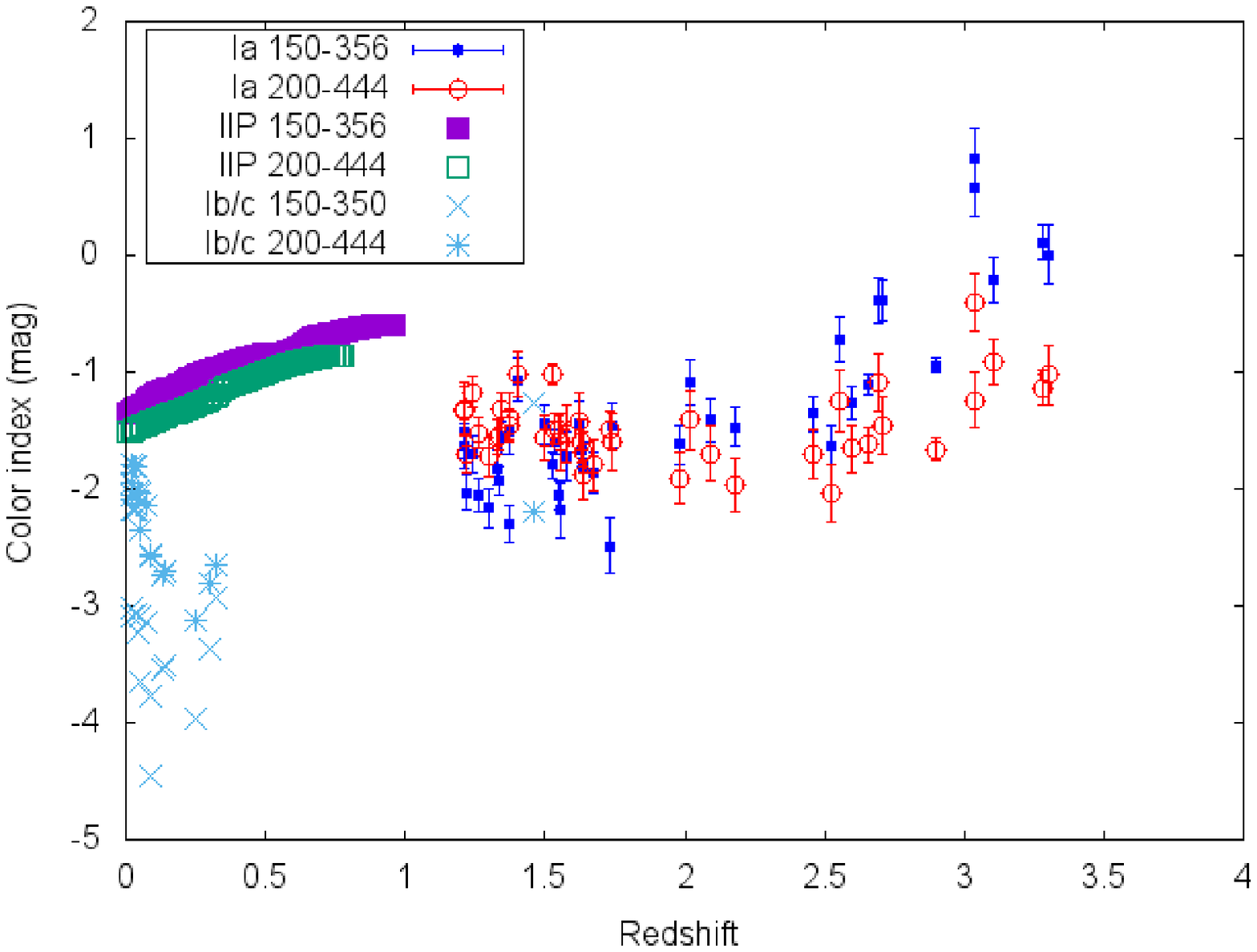}
\caption{Left panel: The position of SNe Ia (circles), Ib/c (asterisks), II-P (squares) and SLSNe (triangles) on the {\it JWST} color-color (F200W$-$F444W vs. F150W$-$F356W) diagram. Right panel: the redshift dependence of the {\it JWST} color indices. Type Ia SNe are shown as circles, while the squares, crosses and asterisks indicate the low-redshift Type II-P and Ib/c SNe that may contaminate the observed sample.}
\label{fig:color1}
\end{figure}

\citet{tanaka13} showed that a near-IR color-color diagram can be a useful tool to identify SLSNe and separate them from fainter foreground transients, like Type II-P SNe that have similar light variation timescales. They proposed the usage of {\it} F200W, F227W and F356W filters in the following combinations: F200W$-$F277W versus F277W$-$F356W ([2.0]$-$[2.8] and [2.8]$-$[3.6] in their notation). They concluded that faint objects that have positive colors ($>0$ magnitude) in both of these combinations are likely to be high-redshift SLSNe.

In the FLARE project \citet{wang17} proposed the application of the F200W vs. F200W$-$F444W color-magnitude diagram for classifying various types of transients to be discovered with {\it JWST} NIRCam. They confirmed that SLSNe indeed occupy a different region than SNe Type Ia or Type II, although they noted that ``ambiguities are inevitable and more data are needed''. 

In this paper we concentrate on identifying $z > 1$ SNe Ia using the NIRCam filters. 
After examining various combinations of the 4 NIRCam filters considered in this paper (F150W, F200W, F356W and F444W, see Section~\ref{sec:survey}), we suggest the following combination: F150W$-$F356W and F200W$-$F444W. A color-color plot with these indices is shown in the left panel of Figure~\ref{fig:color1}. Red open circles represent those Type Ia SNe that passed the strong detection limit in our simulation. The subsample of $z > 2$ SNe is highlighted by blue filled circles. The uncertainty for each color index is computed as during the photo-z estimates (Section~\ref{sec:photoz}).  Also plotted (with squares) are the positions of low-redshift ($z <1$) Type II-P SNe around maximum light. Such low-redshift Type II-P SNe are expected to contaminate the sample of $z > 1$ Type Ia SNe \citep{wang17}. The colors of these Type II-P events are derived using the Nugent-templates \citep{nugent02} after extending the templates up to 5 micron using a Rayleigh-Jeans blackbody tail. Similarly, triangles show the expected colors of SLSNe, whose SEDs are also approximated with blackbodies (see Section~\ref{slsn-highz}) cooling from $T \sim 15,000$ K to $T \sim 6,000$ K as the SLSN evolves from maximum light to $+40$ days post-maximum in rest frame \citep{angus19}. Low-redshift Type Ib/c SNe are also shown as asterisks, even though they are not expected to reach our {\it JWST} detection limit in the F444W filter for $z > 0.5$ redshifts \citep[see Figure~7 in][]{wang17}. Note that the rest-frame mid-infrared SED of Type Ib/c SNe are uncertain, thus, their colors shown here are based on the same Rayleigh-Jeans approximation as used above for the Type II-P SNe.

In the right panel of Figure~\ref{fig:color1} the redshift dependence of the NIRCam color indices are plotted. 
Although using blackbodies is only a rough approximation, the location of $z>2$ SNe Ia in the NIRCam color-color plot seems to be separated from that of core-collapse SNe. This suggests that the classification of SNe based on their NIRCam colors might be feasible. 

\subsection{The Hubble-diagram}
\label{sec:hubble}

\begin{figure}[ht]
\begin{center}
\includegraphics[width=8.5cm]{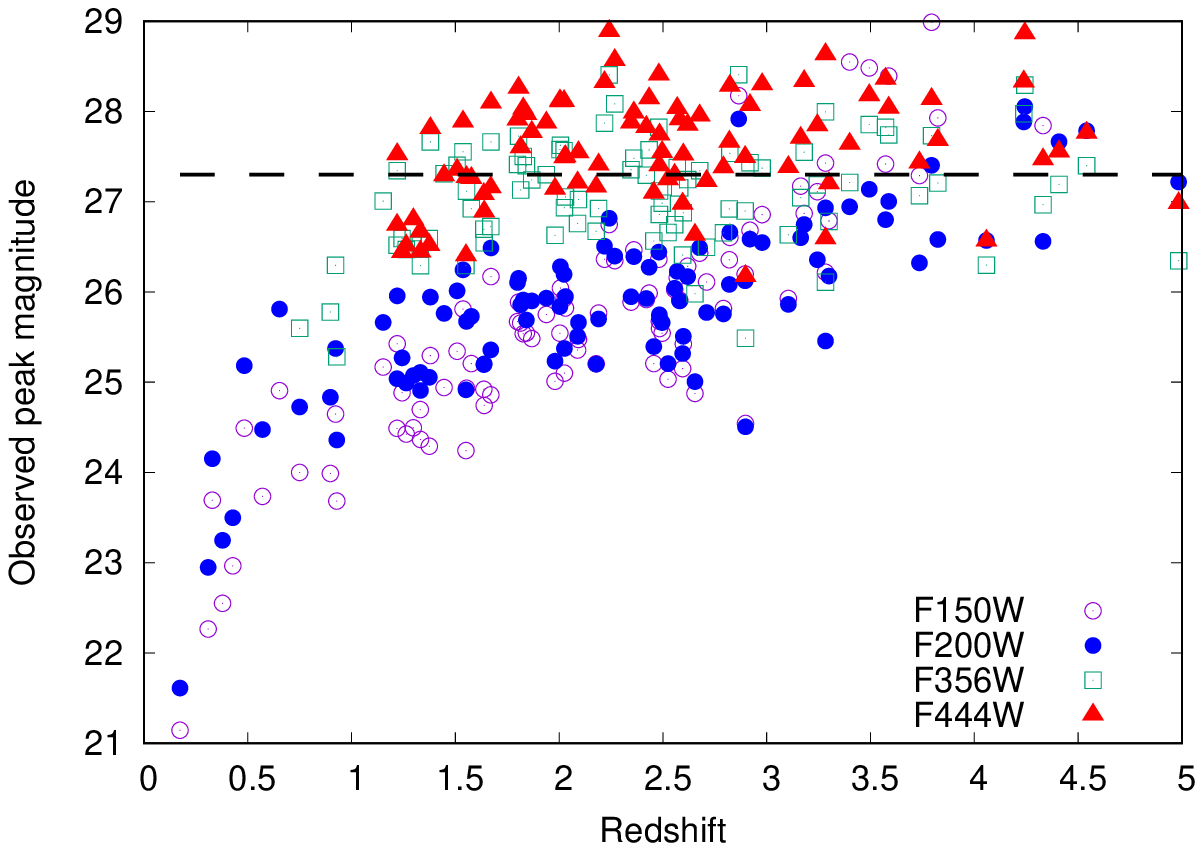}
\includegraphics[width=8.5cm]{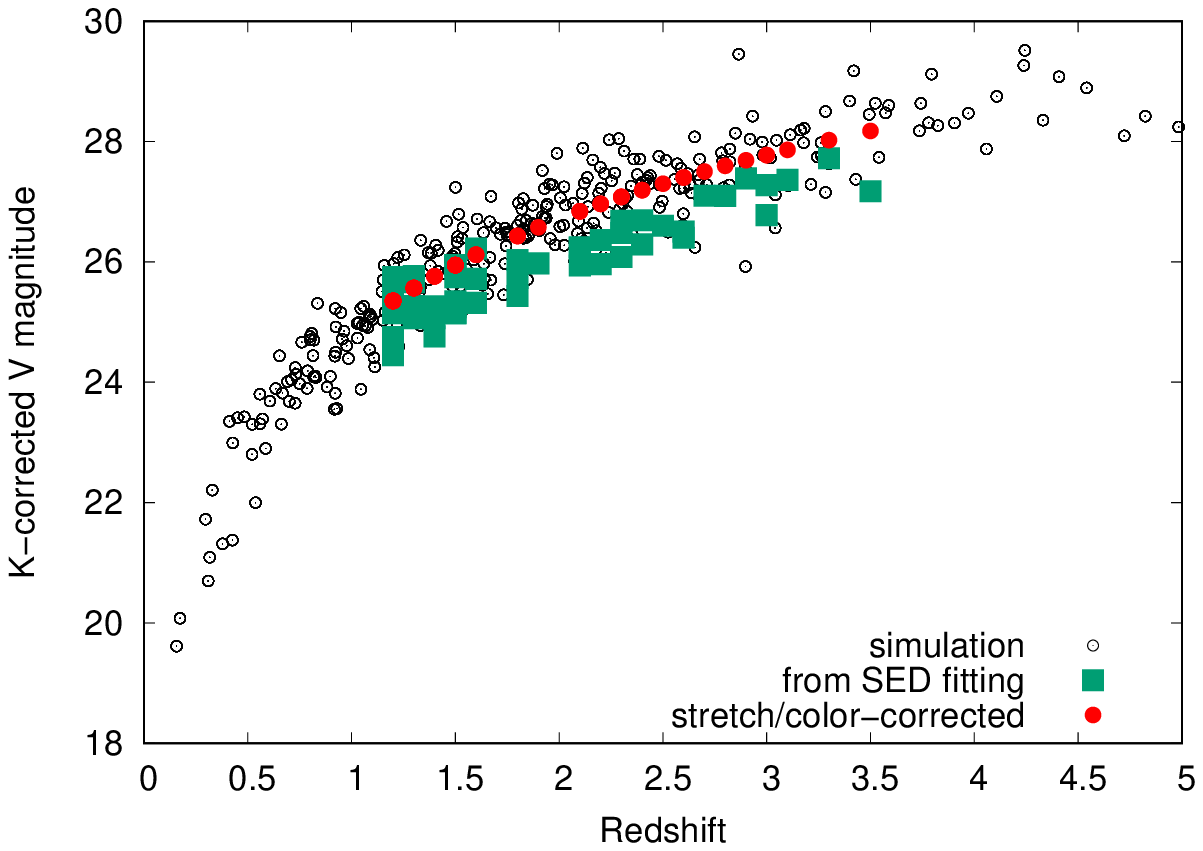}
\caption{Hubble-diagrams of simulated SNe. Left panel: the ``observed'' peak magnitudes in the 4 NIRCam bands (color-coded as indicated in the legend) against redshift. Right-panel: the K-corrected V-band magnitudes of the simulated SNe (black circles), the recovered V-band magnitudes from template fitting (green squares) and
the same data but after correcting for the distribution of the peak brightness as determined from template fitting (red dots). }
\label{fig:hubble}
\end{center}
\end{figure}

The left panel in Figure~\ref{fig:hubble} contains the ``observed'' Hubble-diagram, i.e. the AB-magnitudes of the simulated SNe that passed the detection criterion as functions of redshift, in the 4 NIRCam bands. Dashed horizontal line indicates the assumed detection limit (27.3 AB mag) with NIRCam in the FLARE survey.

As expected, the scatter on this uncorrected Hubble-diagram exceeds 1 mag in all bands above $z \sim 2$ due to i) the random sampling of the light curve in the observer's frame, ii) the intrinsic dispersion in the peak magnitudes of SNe Ia (cf. Figure~\ref{fig:sim-stat}) and iii) K-corrections due to non-negligible redshifts. 

The current state-of-the-art for correcting for all these effects in order to get a ``clean'' Hubble-diagram from SN observations is the application of one of the light curve fitting methods \citep[usually SALT2, see e.g.][]{scolnic18}. In the present case, however, such an approach does not work, as none of the SNe are detected more than twice due to the relatively long adopted cadence (90 days). Thus, alternatives are needed. 

Since the only source of information is the flux in different bands (i.e. the SED of the SN), the peak absolute magnitude of the detected SNe must be determined somehow from their measured SED. This is actually done when measuring the photo-z of the SNe (Section~\ref{sec:photoz}): an output parameter of that fitting is the K-corrected V-band peak absolute magnitude of the best-fit SALT2-template to a particular SN. In the right panel of Figure~\ref{fig:hubble} this quantity is plotted against redshift for all simulated SNe (taken from the input database of the simulation; black circles) as well as their recovered values after template fitting (green squares). The red dots show the case when the correction for the intrinsic distribution of the peak magnitudes (assumed to be Gaussian, Section~\ref{sec:simul}) are also computed via fitting the observed 4-band SEDs with the SALT2 templates as described in Section~\ref{sec:photoz}. This last step would require the knowledge of not only the fiducial peak magnitude of SNe Ia, but also the underlying distribution of the peak magnitudes as a function of redshift, which may not be Gaussian for high-$z$ SNe. Thus, the impressively low scatter of the red dots in the right panel may not be reached from real data without obtaining short-cadence light curves. Since such multi-epoch follow-up observations were quite expensive with {\it JWST}, intense follow-up campaigns with other ground- or space-based instruments (Gemini, Keck, and/or {\it HST}) would be necessary. Designing and optimizing the details of such follow-up campaigns is beyond the scope of the present paper, but it would be an interesting extension to the FLARE survey \citep{wang17}. 

Even if light curves were not available, the reduced scatter of the green squares compared to the $\gtrsim 1$ mag scatter in the left panel of Figure~\ref{fig:hubble} is encouraging. Thus, the SED reconstruction using the extended SALT2 templates seems to be a useful approach in extending the observed Hubble-digram of SNe Ia up to $z \sim 3.5$.

The green data in the right panel in this Figure reveals another effect that may bias the distribution of the measurements on this kind of Hubble-diagram: at $z > 2$ only the SNe that are intrinsically brighter than the mean of their peak brightness distribution are detected. As the red dots suggest, this Malmquist-bias were clearly not present if the correction for the underlying distribution could be made. Nevertheless, this effect must be taken into account when the Hubble-diagram from such high-$z$ SN observations are to be tested with cosmological models. 

Overall, the results above suggests that using such low-cadence {\it JWST} data of SNe Ia for cosmology is not trivial, and more thorough studies, which are beyond the scope of the present paper, are necessary to reach this ambitious goal. 
One of the possibilities is the follow-up of the detected {\it JWST} transients with either ground-, or space-based telescopes, such as {\it Subaru}, {\it VLT} or {\it HST}, as discussed in \citet{wang17}. Details on such a program will be given elsewhere.

\section{Summary}

The results presented in this paper are summarized as follows.

The 90-day cadence survey for transients in the {\it JWST} CVZ, as proposed by the FLARE project \citep{wang17}, is shown to be capable of discovering 5 - 20 SLSNe (depending on the metallicity dependence of their rates), as well as $\sim 50$ SNe Ia between the $1 < z < 4$ redshift interval during the 3 year-long survey. 
    
Although SLSNe could be detected at $z \sim 10$, their low rates probably prevent the discovery of such events above $z \sim 4$. Successful detection of SLSNe at such high redshifts would provide additional observational constraints on the cosmic SFR.

Similarly, SNe Ia discovered at $z > 2$ may be able to constrain their progenitor scenarios (both the SD and the DD channels) and the fraction of prompt Ia population better than the currently available data.
    
From simulated observations of high-redshift SNe Ia with {\it JWST} NIRCam filters we propose the usage of the F200W$-$F444W vs. F150W$-$F356W color-color diagram to select potential SNe Ia from the observations. These color indices show only weak dependence on the rest-frame phase of the SN around peak, and may also be useful in getting rough estimates for the redshift.
    
We show that photometric redshifts can be obtained purely from measuring accurate fluxes in these four {\it JWST} NIRCam bands by fitting the observations with the extended SALT2-templates. The accuracy of these photo-z estimates  ($\sigma_z \sim 0.25$) depends on redshift: the method works better for $z > 2$ SNe when the peak of the SED is redshifted into the region of the NIRCam bands. Similarly, the accuracy of the SN epochs recovered from SED-fitting is $\pm 5$ days.

The same SED-fitting may also be used to get estimates on the K-corrected peak absolute magnitude of the observed SNe in the V-band, provided the extended SALT2-templates indeed represent the high-$z$ SNe as well as their low-$z$ counterparts. At least this correction is necessary to extend the Hubble-diagram to $z > 2$. The resulting data will be still affected by the Malmquist-bias. In order to correct for such effects one would probably need higher cadence light curves, either with {\it JWST} or other ground- and/or space-based telescopes, and more thorough studies are necessary before using these high-redshift SNe Ia observations for cosmology.

\acknowledgments
This work has been supported by the project "Transient Astrophysical Objects"  GINOP 2.3.2-15-2016-00033  of the National Research, 
Development and Innovation Office (NKFIH), Hungary, funded by the European Union.

We are indebted to Lifan Wang, Jeremy Mould, David Rubin, Saul Perlmutter, J. Craig Wheeler, Avishay Gal-Yam, Peter Brown and all other members of the FLARE Collaboration for many hours of enlightening discussion on discovering the high-redshift Universe with {\it JWST} during the preparatory work of the FLARE project. An anonymous referee provided many useful comments and suggestions that led to a significant improvement of this paper. His/her thorough work is gratefully acknowledged. 

\vspace{5mm}
\facilities{JWST}

\software{astropy \citep{2013A&A...558A..33A},  
          sncosmo \citep{2016ascl.soft11017B}
          }


\begin{thebibliography}{}

\bibitem[Angus et al.(2018)]{angus19} Angus, C.~R., Smith, M., Sullivan, M., et al.\ 2018, arXiv:1812.04071 

\bibitem[Astropy Collaboration(2013)]{2013A&A...558A..33A} Astropy Collaboration, Robitaille, T.~P., Tollerud, E.~J., et al.\ 2013, \aap, 558, A33 

\bibitem[Barbary(2011)]{barbaryphd} Barbary, K.~H.\ 2011, Ph.D.~Thesis, University of California, Berkeley 

\bibitem[Barbary et al.(2016)]{2016ascl.soft11017B} Barbary, K., Barclay, T., Biswas, R., et al.\ 2016, Astrophysics Source Code Library, ascl:1611.017 

\bibitem[Bazin et al.(2009)]{bazin09} Bazin, G. et al. 2009, A\&A 499, 653 

\bibitem[Behroozi, Wechsler \& Conroy (2013)]{behroozi13} Behroozi, P. S., Wechsler, R. H., Conroy, C.\ 2013, \apj 770, 57 

\bibitem[Bernstein et al.(2012)]{bernstein12} Bernstein, J.~P., Kessler, R., Kuhlmann, S., et al.\ 2012, \apj, 753, 152 

\bibitem[Betoule et al.(2014)]{betoule14} Betoule, M., Kessler, R., Guy, J., et al.\ 2014, \aap, 568, A22 

\bibitem[Bouwens et al.(2014)]{bouwens14}Bouwens, R.J. et al. 2014, \apj 795, 126

\bibitem[Chatzopoulos et al.(2012)]{manos12} Chatzopoulos, E., Wheeler, J.~C., \& Vinko, J.\ 2012, \apj, 746, 121 

\bibitem[Chatzopoulos et al.(2013)]{manos13} Chatzopoulos, E., Wheeler, J.~C., Vinko, J., Horvath, Z.~L., \& Nagy, A.\ 2013, \apj, 773, 76 

\bibitem[Cooke et al.(2012)]{cooke12} Cooke, J. et al. 2012, Nature 491, 228

\bibitem[Gal-Yam et al.(2009)]{galyam09} Gal-Yam, A., Mazzali, P., Ofek, E.~O., et al.\ 2009, \nat, 462, 624 

\bibitem[Gal-Yam(2012)]{galyam12} Gal-Yam, A.\ 2012, Science, 337, 927 

\bibitem[Graur et al.(2011)]{graur11} Graur, O., Poznanski, D., Maoz, D., et al.\ 2011, \mnras, 417, 916 

\bibitem[Graur et al.(2014)]{graur14} Graur, O., Rodney, S.~A., Maoz, D., et al.\ 2014, \apj, 783, 28 

\bibitem[Grogin et al.(2011)]{grogin11} Grogin, N.~A., Kocevski, D.~D., Faber, S.~M., et al.\ 2011, \apjs, 197, 35 

\bibitem[Guy et al.(2007)]{2007A&A...466...11G} Guy, J., Astier, P., Baumont, S., et al.\ 2007, \aap, 466, 11 

\bibitem[Guy et al.(2010)]{guy10} Guy, J., Sullivan, M., Conley, A., et al.\ 2010, \aap, 523, A7 

\bibitem[Hounsell et al.(2017)]{houn17} Hounsell, R., Scolnic, D., Foley, R.~J., et al.\ 2017, arXiv:1702.01747 

\bibitem[Hopkins \& Beacom(2006)]{hopkins06} Hopkins, A.M., Beacom, B.F. 2006, ApJ 651, 142

\bibitem[Hsiao et al.(2007)]{2007ApJ...663.1187H} Hsiao, E.~Y., Conley, A., Howell, D.~A., et al.\ 2007, \apj, 663, 1187 


\bibitem[Jansen \& Webb Medium Deep Fields IDS GTO Team(2017)]{jansen17} Jansen, R.~A., \& Webb Medium Deep Fields IDS GTO Team, t.~N.-V.~T., and the NEPTDS-Chandra Team 2017, American Astronomical Society Meeting Abstracts \#230, 230, 216.02 

\bibitem[Jones et al.(2013)]{jones13} Jones, D.~O., Rodney, S.~A., Riess, A.~G., et al.\ 2013, \apj, 768, 166 

\bibitem[Kasen \& Bildsten(2010)]{kb10} Kasen, D., \& Bildsten, L.\ 2010, \apj, 717, 245 

\bibitem[Kistler et al.(2009)]{kistler09} Kistler, M.~D., Y{\"u}ksel, H., Beacom, J.~F., Hopkins, A.~M., \& Wyithe, J.~S.~B.\ 2009, \apjl, 705, L104 

\bibitem[Koekemoer et al.(2011)]{koeke11} Koekemoer, A.~M., Faber, S.~M., Ferguson, H.~C., et al.\ 2011, \apjs, 197, 36 

\bibitem[Langer \& Norman(2006)]{ln06} Langer, N., Norman, C. A.\ 2006, \apj 638, L63

\bibitem[Leloudas et al.(2015)]{leloudas15} Leloudas, G., Schulze, S., Kr{\"u}hler, T., et al.\ 2015, \mnras, 449, 917 

\bibitem[Livio \& Mazzali(2018)]{livio18} Livio, M., \& Mazzali, P.\ 2018, \physrep, 736, 1 

\bibitem[Lunnan et al.(2014)]{lunnan14} Lunnan, R., Chornock, R., Berger, E., et al.\ 2014, \apj, 787, 138 


\bibitem[Lunnan et al.(2015)]{lunnan15} Lunnan, R., Chornock, R., Berger, E., et al.\ 2015, \apj, 804, 90 

\bibitem[Madau \& Dickinson(2014)]{madi14} Madau, P., \& Dickinson, M.\ 2014, \araa, 52, 415 

\bibitem[Maoz et al.(2014)]{maoz14} Maoz, D., Mannucci, F., \& Nelemans, G.\ 2014, \araa, 52, 107 

\bibitem[Moe \& Di Stefano (2013)]{moe13} Moe, M., Di Stefano, R. 2013, Binary Paths to Type Ia Supernovae Explosions, Proceedings of the 281st IAU Symposium, Cambridge University Press, Volume 281, p. 240

\bibitem[Moriya et al.(2018)]{moriya18} Moriya, T.~J., Tanaka, M., Yasuda, N., et al.\ 2018, arXiv:1801.08240 

\bibitem[Mutch et al.(2016)]{mutch16} Mutch, S.~J., Geil, P.~M., Poole, G.~B., et al.\ 2016, \mnras, 462, 250 

\bibitem[NEPTDS(2017)]{neptsd17} NEPTDS-Chandra Team, 2017, AAS Meeting Abstracts Vol. 230, 216.02

\bibitem[Neill et al.(2011)]{neill11} Neill, J.~D., Sullivan, M., Gal-Yam, A., et al.\ 2011, \apj, 727, 15 

\bibitem[Nelemans et al.(2013)]{nelemans13} Nelemans G., Toonen, S., Bours, M. 2013, Binary Paths to Type Ia Supernovae Explosions, Proceedings of IAUS 281, Cambridge University Press, Volume 281, p. 225

\bibitem[Nicholl et al.(2017)]{nicholl17} Nicholl, M., Guillochon, J., \& Berger, E.\ 2017, \apj, 850, 55 

\bibitem[Nugent et al.(2002)]{nugent02} Nugent, P., Kim, A., \& Perlmutter, S.\ 2002, \pasp, 114, 803 

\bibitem[Oesch et al.(2015)]{oesch15} Oesch, P. A. et al. 2015, \apj 808, 104

\bibitem[Oesch et al.(2018)]{oesch18} Oesch, P.~A., Bouwens, R.~J., Illingworth, G.~D., Labb{\'e}, I., \& Stefanon, M.\ 2018, \apj, 855, 105 

\bibitem[Planck collaboration(2014)]{planck14} Planck collaboration, 2014, A\&A 571, 16 

\bibitem[Planck Collaboration et al.(2018)]{planck18} Planck Collaboration, Aghanim, N., Akrami, Y., et al.\ 2018, arXiv:1807.06209 

\bibitem[Perley et al.(2016)]{perley16} Perley, D.~A., Quimby, R.~M., Yan, L., et al.\ 2016, \apj, 830, 13 

\bibitem[Postman et al.(2012)]{postman12} Postman, M., Coe, D., Ben{\'{\i}}tez, N., et al.\ 2012, \apjs, 199, 25 

\bibitem[Prajs et al.(2017)]{prajs17} Prajs, S., Sullivan, M., Smith, M., et al.\ 2017, \mnras, 464, 3568 

\bibitem[Quimby et al.(2011)]{quimby11} Quimby, R.~M., Kulkarni, S.~R., Kasliwal, M.~M., et al.\ 2011, \nat, 474, 487 

\bibitem[Quimby et al.(2013)]{quimby13} Quimby, R. et al. 2013, \mnras 431, 912

\bibitem[Raskin et al.(2009)]{raskin09} Raskin, C., Scannapieco, E., Rhoads, J., \& Della Valle, M.\ 2009, \apj, 707, 74 

\bibitem[Reg\H os (2013)]{regos13} Reg\H os 2013, Binary Paths to Type Ia Supernovae Explosions, Proceedings of IAUS 281, Cambridge University Press, Volume 281, p. 26

\bibitem[Richardson et al.(2014)]{2014AJ....147..118R} Richardson, D., Jenkins, R.~L., III, Wright, J., \& Maddox, L.\ 2014, \aj, 147, 118 

\bibitem[Riess et al.(2018)]{riess18} Riess, A.~G., Rodney, S.~A., Scolnic, D.~M., et al.\ 2018, \apj, 853, 126 

\bibitem[Robertson \& Ellis(2012)]{re12} Robertson, B.~E., \& Ellis, R.~S.\ 2012, \apj, 744, 95 

\bibitem[Rodney et al.(2014)]{rodney14} Rodney, S.~A., Riess, A.~G., Strolger, L.-G., et al.\ 2014, \aj, 148, 13 

\bibitem[Rodney et al.(2015)]{rodney15} Rodney, S.~A., Riess, A.~G., Scolnic, D.~M., et al.\ 2015, \aj, 150, 156 

\bibitem[Rubin et al.(2018)]{rubin18} Rubin, D., Hayden, B., Huang, X., et al.\ 2018, \apj, 866, 65 

\bibitem[S{\'a}nchez et al.(2014)]{sanchez14} S{\'a}nchez, C., Carrasco Kind, M., Lin, H., et al.\ 2014, \mnras, 445, 1482 

\bibitem[Scannapieco \& Bildsten(2005)]{scanna05} Scannapieco, E., \& Bildsten, L.\ 2005, \apjl, 629, L85 

\bibitem[Scolnic \& Kessler(2016)]{scolnic16} Scolnic, D., \& Kessler, R.\ 2016, \apjl, 822, L35 

\bibitem[Scolnic et al.(2018)]{scolnic18} Scolnic, D.~M., Jones, D.~O., Rest, A., et al.\ 2018, \apj, 859, 101 

\bibitem[Strolger et al.(2004)]{strolger04} Strolger, L-G. et al. 2004, \apj 613, 200

\bibitem[Tanaka et al.(2013)]{tanaka13} Tanaka, M., Moriya, T. J., Yoshida, N. 2013, \mnras 435, 2483

\bibitem[Trenti et al.(2013)]{trenti13} Trenti, M., Perna, R., \& Tacchella, S.\ 2013, \apjl, 773, L22 

\bibitem[Virgili et al.(2011)]{virgili11} Virgili, F.~J., Zhang, B., Nagamine, K., \& Choi, J.-H.\ 2011, \mnras, 417, 3025 

\bibitem[Wang et al.(2017)]{wang17} Wang, L. et al., 2017 arXiv:1710.07005

\bibitem[Wang(2007)]{wang07} Wang, Y.\ 2007, \apjl, 654, L123 

\bibitem[Wang et al.(2007)]{wangetal07} Wang, Y., Narayan, G., \& Wood-Vasey, M.\ 2007, \mnras, 382, 377 

\bibitem[Woosley \& Bloom(2006)]{wb06} Woosley, S.~E., \& Bloom, J.~S.\ 2006, \araa, 44, 507 

\bibitem[Yoon et al.(2006)]{yoon06} Yoon, S.-C., Langer, N., \& Norman, C.\ 2006, \aap, 460, 199 

\bibitem[Yuan et al.(2010)]{yuan10} Yuan, F., Quimby, R.~M., Wheeler, J.~C., et al.\ 2010, \apj, 715, 1338 

\end{thebibliography}
\end{document}